\documentclass[american,aps,pra,reprint,floatfix,nofootinbib,superscriptaddress]{revtex4-1}
\usepackage[unicode=true,pdfusetitle, bookmarks=true,bookmarksnumbered=false,bookmarksopen=false, breaklinks=false,pdfborder={0 0 0},backref=false,colorlinks=false]{hyperref}
\hypersetup{colorlinks,linkcolor=blue,citecolor=red,urlcolor=cyan}
\usepackage{graphics,epstopdf,amsthm,amsmath,amssymb,mathptmx,colortbl,color,bm,framed,mathrsfs}
\usepackage[T1]{fontenc}
\usepackage[up]{subfigure}
\usepackage[braket, qm]{qcircuit}
\usepackage{commath}
\usepackage{float}

\makeatother

\usepackage[demo]{graphicx}

\usepackage[normalem]{ulem}

\newcommand{\bes}{\begin{subequations}}
\newcommand{\ees}{\end{subequations}}

\begin{document}
\title{Demonstration of fidelity improvement using dynamical decoupling with superconducting qubits} 

\author{Bibek Pokharel}
\affiliation{Department of Electrical Engineering, University of Southern California, Los Angeles, CA 90089}

\author{Namit Anand}
\affiliation{Department of Physics and Astronomy, University of Southern California, Los Angeles, CA 90089}

\author{Benjamin Fortman}
\affiliation{Department of Chemistry, University of Southern California, Los Angeles, CA 90089}

\author{Daniel A. Lidar}
\affiliation{Department of Electrical Engineering, University of Southern California, Los Angeles, CA 90089}
\affiliation{Department of Physics and Astronomy, University of Southern California, Los Angeles, CA 90089}
\affiliation{Department of Chemistry, University of Southern California, Los Angeles, CA 90089}
\affiliation{Center for Quantum Information Science \& Technology, University of Southern California, Los Angeles, CA 90089, USA}

\date{\today}

\begin{abstract}
Quantum computers must be able to function in the presence of decoherence. The simplest strategy for decoherence reduction is dynamical decoupling (DD), which requires no encoding overhead and works by converting quantum gates into decoupling pulses. Here, using the IBM and Rigetti platforms, we demonstrate that the DD method is suitable for implementation in today's relatively noisy and small-scale cloud-based quantum computers. Using DD, we achieve substantial fidelity gains relative to unprotected, free evolution of individual superconducting transmon qubits. To a lesser degree, DD is also capable of protecting entangled two-qubit states. We show that dephasing and spontaneous emission errors are dominant in these systems, and that different DD sequences are capable of mitigating both effects. Unlike previous work demonstrating the use of quantum error correcting codes on the same platforms, we make no use of post-selection and hence report unconditional fidelity improvements against natural decoherence.
\end{abstract}
\maketitle

\textit{Introduction}.---%
Two decades after the first detailed quantum computing proposals \cite{Cirac:95,Loss:1998zr,Kane:98,Nakamura:99}, rudimentary gate-model quantum computers (QCs) based on superconducting transmon qubits with coherence times in the microseconds range are finally available and remotely accessible via public cloud-based services. Interest in these platforms, made publicly available so far by IBM, Rigetti, and Alibaba,
has been high, and numerous experiments have been reported demonstrating a variety of quantum protocols~\cite{Alsina:2016aa,berta2016entropic,ibm2017hardy,ibm2017envariance} and algorithms~\cite{Riste:2017aa,ibm2017adder,Coles:2018aa}.
Given their present intermediate scale of $10$-$20$ fairly noisy qubits, gates, and measurements~\cite{Preskill:2018aa}, the current QCs are particularly very well suited to tests of simple quantum error correction and suppression protocols. Indeed, a variety of quantum error correction (QEC) experiments on cloud based platforms have been reported~\cite{ibm2016simon,2017repetition,ibm2017useful,2017cpc,2017simpleqec,2018arXivQFTsmall,Harper:2018aa}. 
However, so far this body of work has not offered a demonstration that QEC can result in improvements for general decoherence while applying standard initialization, gates, and readout operations (we review these studies in Appendix~\ref{app:QEC-review}). The main reason appears to be that the overhead introduced by QEC results in error rates that are too high to be compensated by the schemes that have been tried so far, and claims of improvement have had to resort to cleverly avoiding the execution of actual initialization and key gate operations~\cite{Harper:2018aa}. We note that some QEC successes have been reported in other transmon qubit systems~\cite{2012Natur.482..382R,Ofek:2016aa}.

Here, rather than attempting to demonstrate error correction, we focus on error suppression. Specifically, we seek to mitigate the effects of decoherence using dynamical decoupling (DD)~\cite{Viola:98,Duan:98e,zanardi1999symmetrizing,viola1999dynamical}, one of the simplest strategies available in the toolkit of quantum error mitigation~\cite{lidar2013quantum}. We demonstrate that DD is capable of extending the lifetimes of single-qubit states as well as entangled two-qubit states. To the best of our knowledge, this amounts to the first unequivocal demonstration of successful decoherence mitigation in cloud-based superconducting qubit platforms. Moreover, as a test of the robustness of our results we performed our DD experiments on three of the cloud-based systems, the $16$-qubit IBMQX5, $5$-qubit IBMQX4, and the $19$-qubit Rigetti Acorn chips. Given their similarities they provide suitable platforms for independent tests of the performance of DD, and we expect that the lessons drawn will have wide applicability.

\textit{Dynamical Decoupling}.---%
DD is a well-established method designed to suppress decoherence via the application of pulses applied to the system, that cancel the system-environment interaction to a given order in time-dependent perturbation theory~\cite{Lidar:2014aa}. A large variety of DD protocols has been developed and tested, with some of the more advanced protocols capable of reducing decoherence to arbitrarily low levels under the assumption of perfectly implemented instantaneous pulses with arbitrarily small pulse intervals~\cite{cdd,udd,qdd,Wang:10,Khodjasteh:2010qd}. In reality, pulses are of course never implemented perfectly, have a minimum duration, and pulse intervals are finite. Various specialized DD sequences have been developed to handle such conditions as well~\cite{Souza:2011aa,Souza:2012aa,genetic,Kabytayev:2014aa,Genov:2017aa}, and it has been shown that imperfect DD can improve the performance of fault-tolerant quantum computation~\cite{Ng:2011dn}. Here, as a proof of principle, we explore the benefits of using primarily the simplest DD sequence, designed to offer only first order cancellation and not designed with robustness against pulse imperfections in mind, namely the XY4 ``universal decoupling" sequence~\cite{viola1999dynamical}. This sequence consists of a simple repetition of the pulse pattern $XYXY$, where $X$ and $Y$ are rotations by $\pi$ about the $x$ and $y$ axes of the single-qubit Bloch sphere, and the system evolves freely for time $\tau$ between the pulses. Starting from the system-bath Hamiltonian 
$H = H_S+ H_B +H_{SB}$,
with the interaction term 
$H_{SB} =\sum_{i=1}^N \sum_{\alpha \in\{x,y,z\}} \sigma_i^{\alpha} \otimes B_i^{\alpha}$, where $ \sigma_i^{\alpha}$ and $B_i^{\alpha}$ are, respectively, Pauli matrices acting on qubit $i$, and general operators acting on the bath,
the action of the XY4 sequence is readily shown to result in the elimination of $H_{SB}$ to first order in $\tau$ in the joint system-bath unitary propagator, under the assumption of instantaneous $X$ and $Y$ pulses.

\begin{figure}[h]
\hspace{.5cm}
\Qcircuit @C=1em @R=0.7em @!R {
& \lstick{\ket{0}}& \qw & \gate{U(\theta, \phi,\lambda)} &   \gate{X} & \gate{Y} & \gate{X} & \gate{Y}   & \gate{U^{\dagger}(\theta, \phi,\lambda)} & \measureD{Z}
}
\vspace{.4cm}
\Qcircuit @C=1em @R=0.7em @!R {
& \lstick{\ket{0}}& \qw & \gate{U(\theta, \phi,\lambda)}  &  \gate{I} & \gate{I} & \gate{I} & \gate{I}  &   \gate{U^{\dagger}(\theta, \phi,\lambda)} & \measureD{Z}
}
\caption{Quantum circuit for (top) the DD sequence $XYXY$ and (bottom) the free evolution of the initial state $|\psi\rangle = U(\theta, \phi,\lambda) \ket{0}$, following by a measurement in the computational basis. Only a single repetition is depicted.}
\label{figure:DDvsIonTheta}
\end{figure}
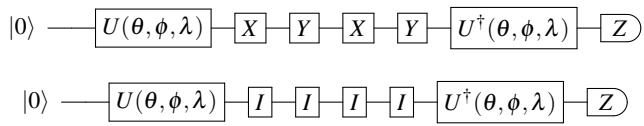

\textit{Methodology.}---%
The native single gates on the IBM and Rigetti platforms are rotations about the $z$ and $x$ axes of the Bloch sphere, $R_{\alpha} (\phi) = \exp[i (\phi/2) \sigma^{\alpha}]$, with $\alpha\in\{y,z\}$ (see Appendix \ref{sub:machinespecs} for more details about these platforms). Arbitrary single-qubit unitaries can be applied by specifying Euler angles $\theta,\phi,\lambda$ such that 
\begin{align}
U(\theta, \phi,\lambda) & =  i R_{z} (\phi) R_{y} (\theta) R_{z} (\lambda) .
\label{eq:angular} 
\end{align} 
Since DD is expected to provide quantum error suppression for arbitrary initial states, we tested the performance of DD on a variety of initial states by repeatedly preparing single-qubit states of the form 
$\ket{\psi} = U(\theta, \phi,\lambda) \ket{0}$,
where $\ket{0}$ and $\ket{1}$ are computational basis states (eigenstates of $\sigma^z$).
It should be noted that in transmon qubits the $\ket{0}$ and $\ket{1}$ states are, respectively, the ground and first excited states; this has important implications as discussed below.
For each single-qubit state $\ket{\psi}$, we performed two sets of experiments: one under the action of DD, and another with free evolution. DD pulses were applied as the gates $X=i \exp[- i(\pi/2)\sigma^x] = U(\pi,0,\pi)$ and $Y=i \exp[-i(\pi/2)\sigma^y] = U(\pi,2\pi,0)$. On the IBMQX5 (Acorn) chip each single-qubit pulse lasted $80$ns ($40$ns), with a $10$ns buffer of free evolution between pulses, and each such run was repeated $8192$ ($1000$) times. Identity pulses were implemented as free evolutions lasting $90$ns ($50$ns) on the IBMQX5 (Acorn) chip. 
Since measurements in bases other than $Z$ were not possible, we applied $U^{\dagger}(\theta, \phi,\lambda)$ at the very end of each run and measured the final state of each qubit in the $Z$ basis. A sample circuit showing the state preparation, evolution under a single repetition of DD, and measurement is shown in Fig.~\ref{figure:DDvsIonTheta}. 

Our key performance metric is the fidelity between the input and the output states, defined as the total number of $\ket{0}$ states empirically observed divided by the total number of repetitions. 
We considered two types of initial conditions. In ``type 1", $\theta$ was varied in $16$ equidistant steps in the range $[0,\pi]$, with $\phi = \lambda = 0$. 
This corresponds to a sequence of states (superpositions for $0<\theta<\pi$) of the form $\ket{\psi} = \cos \left( {\theta}/{2} \right) \ket{0} + \sin \left( {\theta}/{2} \right) \ket{1}$ (up to a global phase).
In ``type 2", we considered a set of $30$ random initial conditions sampled uniformly from the Bloch sphere along with the $6$ eigenstates of the Pauli matrices, i.e., $\ket{0}, \ket{1}, \ket{\pm} = \frac{1}{\sqrt{2}} \left( \ket{0} \pm \ket{1} \right), \ket{\pm i} =\frac{1}{\sqrt{2}} \left( \ket{0} \pm i \ket{1} \right)$. 

\begin{figure}[t]
\raggedright
\vspace{-2cm}
\includegraphics[width = .48\textwidth]{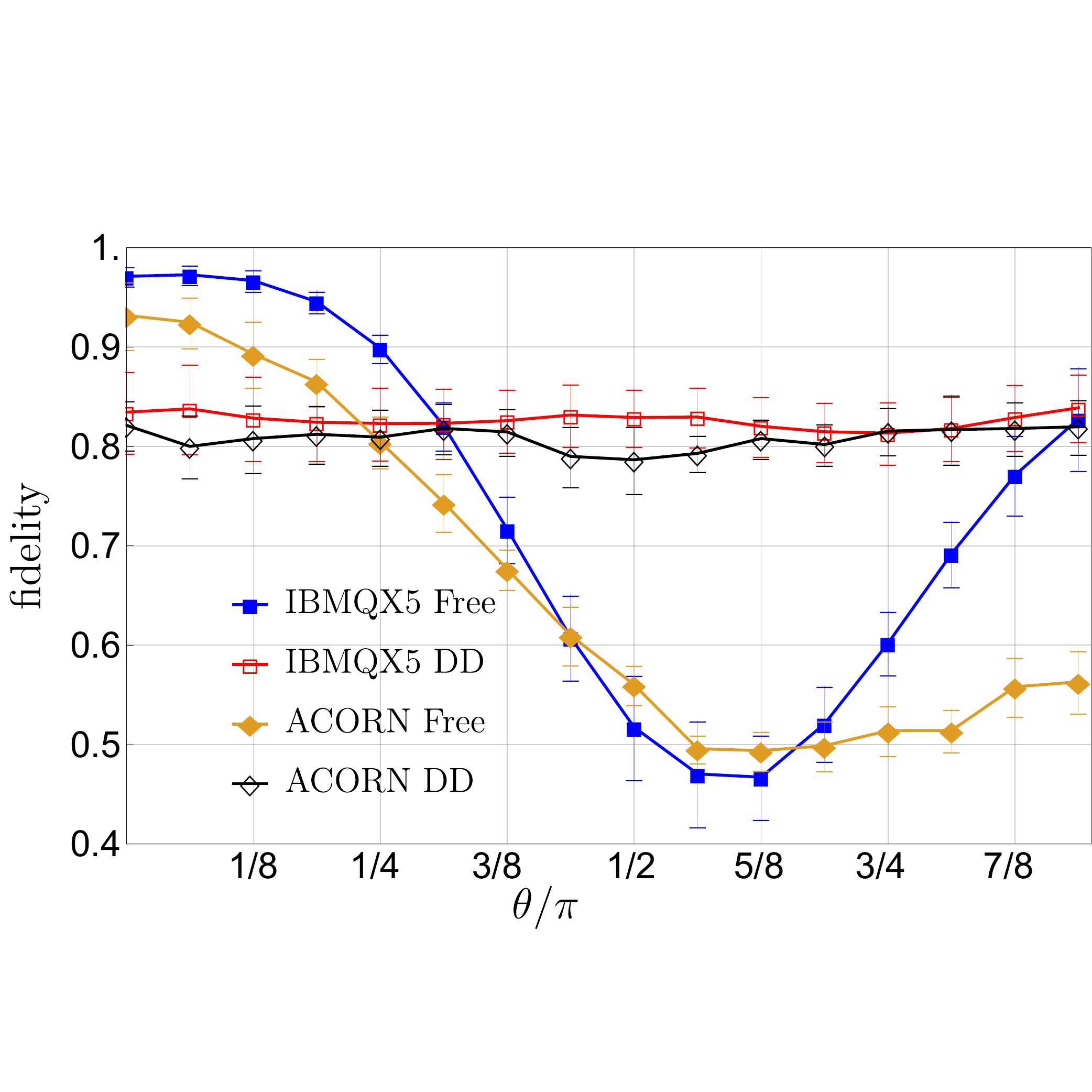}
\vspace{-2cm}
\caption[ ]{(Color online) Mean fidelity over $16$ qubits of IBMQX5 and $15$ qubits of Acorn, for initial states $\ket{\psi} = -i[\cos \left( {\theta}/{2} \right) \ket{0} + \sin \left( {\theta}/{2} \right) \ket{1}]$. Results shown are under DD using XY4 and under free evolution. IBMQX5: after $N=40$ pulses, i.e., $10$ repetitions of the base XY4 sequence. DD improves the fidelity only for states with $\theta \gtrsim \pi/3$. Acorn: after $N=192$ pulses, i.e., $48$ repetitions of the base XY4 sequence. DD improves the fidelity only for states with $\theta \gtrsim \pi/4$. Throughout we report $2 \sigma$ error bars ($95\%$ confidence intervals) calculated using the bootstrap method (for more details see Appendix~\ref{sub:boot}).} 
\label{fig:angDep}
\end{figure}

\textit{Single-qubit results.}---%
We first tested the dependence on the initial state using type 1 preparation. The results are shown in Fig.~\ref{fig:angDep}, which displays the fidelity as a function of $\theta$ for the IBMQX5 and Acorn after $40$ and $192$ pulses, respectively. Under free evolution, the fidelity is relatively high for $\theta \approx 0$ (corresponding to the ground state $\ket{0}$) on both devices, and approaches a clear minimum for $\theta\approx\frac{5\pi}{8}$, i.e., a superposition state slightly biased towards $\ket{1}$. On both devices the free evolution fidelity rises towards the excited state $\ket{1}$, but remains well below that of the ground state. Thus coherent superposition states undergo significant dephasing and the excited state $\ket{1}$ undergoes spontaneous emission (SE) and relaxes to the ground state. 

The situation is dramatically different under DD. As Fig.~\ref{fig:angDep} clearly demonstrates, on both devices the $\theta$-dependence is essentially eliminated. It is clear that the overall fidelity (averaged over $\theta$) increases significantly, while DD reduces the fidelity of states close to the ground state. This is not surprising, given that the XY4 sequence is designed to suppress all single-qubit error types equally. We provide a more detailed, instance-dependent analysis of the initial state dependence in Appendix~\ref{app:initstate}.

\begin{figure}[t]
\raggedright
\includegraphics[width = .5\textwidth]{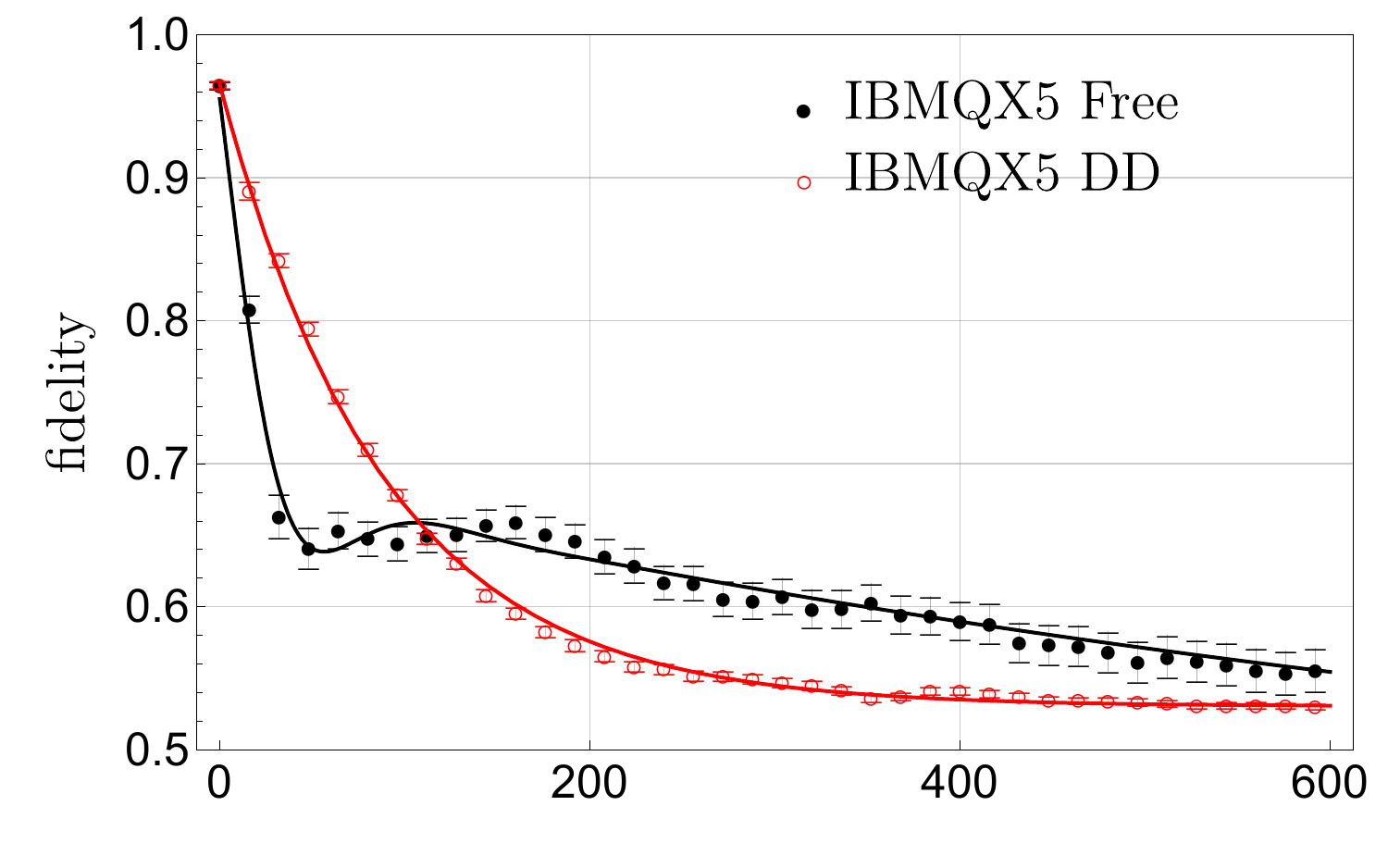}
\includegraphics[width = .5\textwidth]{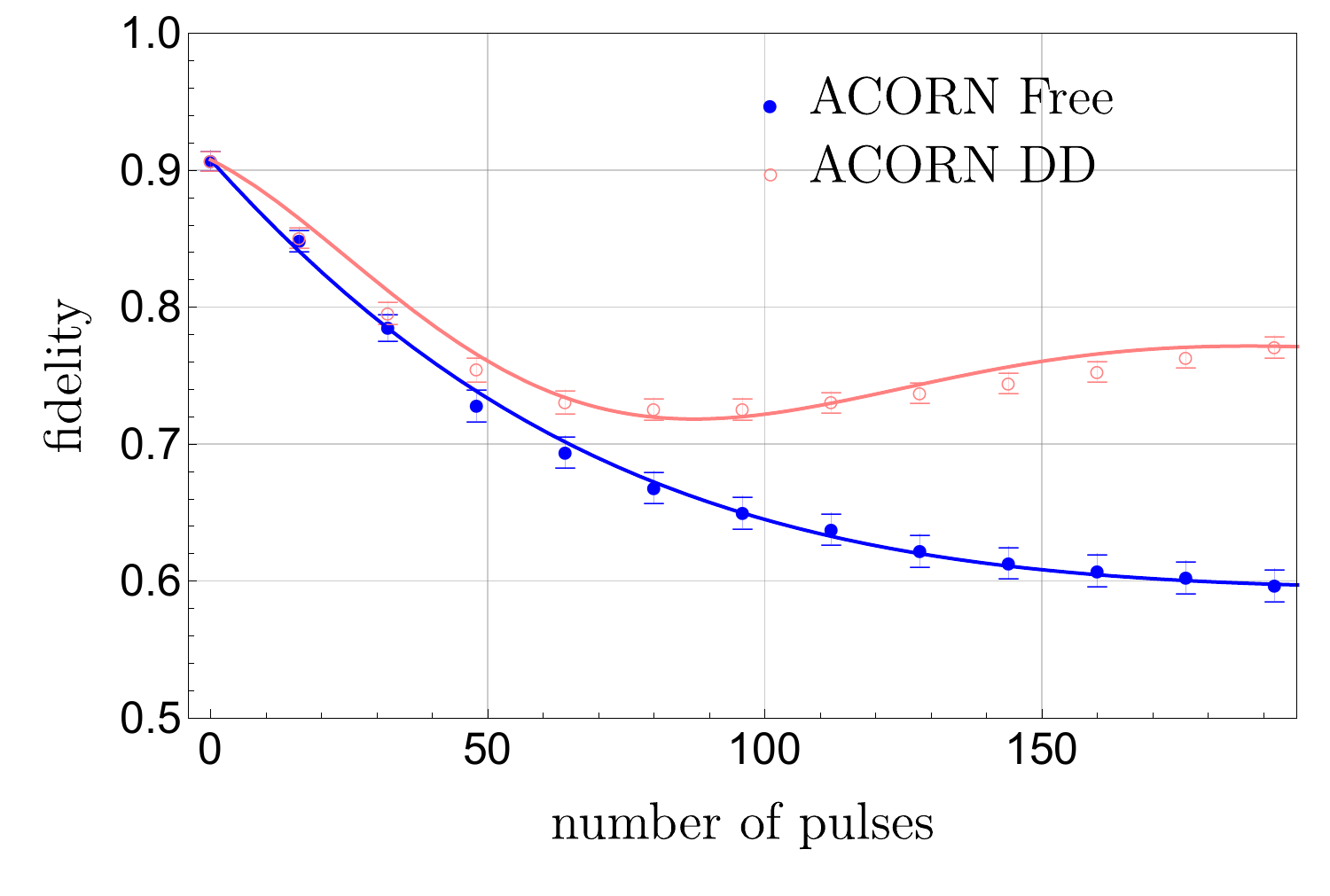}
\caption[ ]{(Color online) 
Mean fidelity, after averaging over all qubits, and all $36$ initial conditions in type 2 preparation, as a function of the number of pulses. The pulse interval is the shortest possible: $90$ns for IBMQX5 (top), $50$ns for Acorn (bottom). Solid lines are fits to Eq.~\eqref{eq:fit}, with fit parameters as per Table~\ref{tab:singleQubitFidelities}.} 
  \label{fig:timeEvolAvg}
\end{figure}

Next, we tested performance under type 2 preparation. The results are shown in Fig.~\ref{fig:timeEvolAvg}.
%
For IBMQX5, DD significantly reduces the fidelity decay up to $N \approx 110$ pulses. The free evolution fidelity decays rapidly but has a shallow minimum at $N\approx 60$, then surpasses the fidelity under DD for $N>110$, which continues to decay exponentially. This exponential decay is consistent with Markovian dynamics. 

The situation is rather different for Acorn. First, we note that the initial fidelity (determined by the initialization and readout errors) is lower for Acorn than for IBMQX5: $\sim0.91$ and $\sim0.96$, respectively. Second, the fidelity under DD is consistently greater than under free evolution, and the roles are reversed: free evolution is very nearly Markovian (exponential decay) while under DD it exhibits a recurrence. These fidelity differences suggest that the environments are different for the two QCs, with the native IBMQX5 environment being non-Markovian, while that of Acorn is more Markovian. Conversely, DD removes the non-Markovian component for IBMQX5, while it introduces a non-Markovian component for Acorn. A detailed study of these effects is beyond the scope of this work, though we may speculate that the non-Markovianity is due to residual low-frequency noise (e.g., $1/f$) in the IBMQX5 case, and that the DD pulses themselves introduce low frequency noise in the Acorn case.

To quantify the fidelity decay with and without DD we fit the data to a modulated exponential decay with three free parameters $\lambda, \alpha$ (dimensionless decay times) and $\gamma$ (dimensionless modulation frequency):
\bes
\begin{align}
  F(N) &= c f(N) + c_{0}, \quad f(N) = e^{- N/\lambda} \cos (N \gamma) + e^{- N/\alpha}.\\
  c &= \frac{F_{N_{max}} - F_{0}}{f(N_{max}) -1} , \quad c_0 = F_{0}-c .
\end{align}
  \label{eq:fit}
\ees
Here $F_0$ is the initial fidelity, $F_{N_{max}}$ is the fidelity at $N=592$ $(192)$ for IBMQX5 (Acorn). The deviation of $F_{N_{max}}$ from $1$ accounts for the initialization errors, readout errors, and decoherence that were not cancelled by DD, as well as the errors accumulated during the application of the imperfect DD pulses, arising from imperfect control over the pulse shape, duration, and interval. Table~\ref{tab:singleQubitFidelities} summarizes the values of the fit parameters. 

\begin{table}
\centering
\resizebox{\columnwidth}{!}{
 \begin{tabular}{|| c c c c c c c||}
 
 \hline
 Machine & Evolution & $F_0 \times 10^{-2}$ &$F_{N_{max}}  \times 10^{-2}$ & $\lambda$ & $\alpha$ &$\gamma$  \\
 \hline\hline
 IBMQX5 & Free       & $96.5 \pm 0.1$ & $55.6 \pm  0.7$ & $28.9 \pm 1.2$ & $910 \pm 5$ & $0.73 \pm 0.12$ \\ \hline
 IBMQX5 & DD         & $96.5 \pm 0.1$ & $53.1 \pm  0.1$ & $88.4 \pm 0.3$ & $\infty$    & $0$             \\ \hline
 Acorn  & Free & $90.8 \pm 0.4$ & $59.8 \pm  0.6$ & $68.1 \pm 1.3$ & $\infty$    & $0.14 \pm 0.11$ \\ \hline
 Acorn  & DD  & $90.8 \pm 0.4$ & $77.1 \pm  0.4$ & $74.9 \pm 0.9$ & $\infty$    & $0.50 \pm .03 $ \\ \hline
 \hline
\end{tabular}
}
\caption{Fit parameters when Eq.~\eqref{eq:fit} is used to fit the mean fidelities in Fig.~\ref{fig:timeEvolAvg}. The first decay constant, $\lambda$, is significantly increased under DD. The second decay constant, $\alpha$, is effectively infinite for all evolutions other than IBMQX5's free evolution. The modulation frequency $\gamma$ vanishes for IBMQX5 under DD and is near zero for Acorn under free evolution, consistent with purely exponential fidelity decay, i.e., Markovian evolution.}
\label{tab:singleQubitFidelities}
\end{table}

While $\lambda$ quantifies the sharp decay during the beginning of the evolution, evolution at longer timescales is quantified by $\alpha$. In the relatively short depth circuits which are our focus here, the role of $\lambda$ is dominant. The most significant finding for IBMQX5 is that the initial decay time characterized by $\lambda$ is more than tripled in the presence of DD. While the improvement in decay time is much more modest for Acorn, the result is a sense even better than for IBMQX5, in that DD improves its fidelity for \emph{all} $N$ we were able to test. We also tested DD on the $5$-qubit IBMQX4, with similar results (see Appendix~\ref{sub:machinespecs} for details).

\textit{Dephasing vs. spontaneous emission}.---%
Figure~\ref{fig:angDep} shows that both dephasing and SE play important roles. This is studied  in more detail in Appendix~\ref{app:initstate}, where we show that for initial states close to the ground state $\ket{0}$, DD is worse than free evolution, but for superposition states susceptible to dephasing and states close to the excited state $\ket{1}$ susceptible to SE, there is a clear benefit in using the XY4 sequence. In light of this, it is interesting to try to address one of these error sources at a time. A DD sequence that suppresses only dephasing ($\sigma^z$) errors is $(XI)^N$ or $(YI)^N$ ($N$ repetitions of $XI$ or $YI$), since $X$ and $Y$ anticommute with $\sigma^z$. Likewise, SE is suppressed by $(ZI)^N$, since $Z$ anticommutes with $\sigma^-$. We report on results for these sequences in Appendix~\ref{app:SE}; they underperform XY4, as expected, but both lead to a substantial slowing down of fidelity decay, with dephasing suppression being the dominant effect, accounting for nearly $90\%$ of the value of $\lambda$ under XY4. This can be viewed as an example of using DD as a diagnostic tool, to identify the relative dominance of different decoherence channels~\cite{Byl11a,Alvarez:2011aa}.

\begin{figure}[t]
\raggedright
  \includegraphics[width = .48\textwidth]{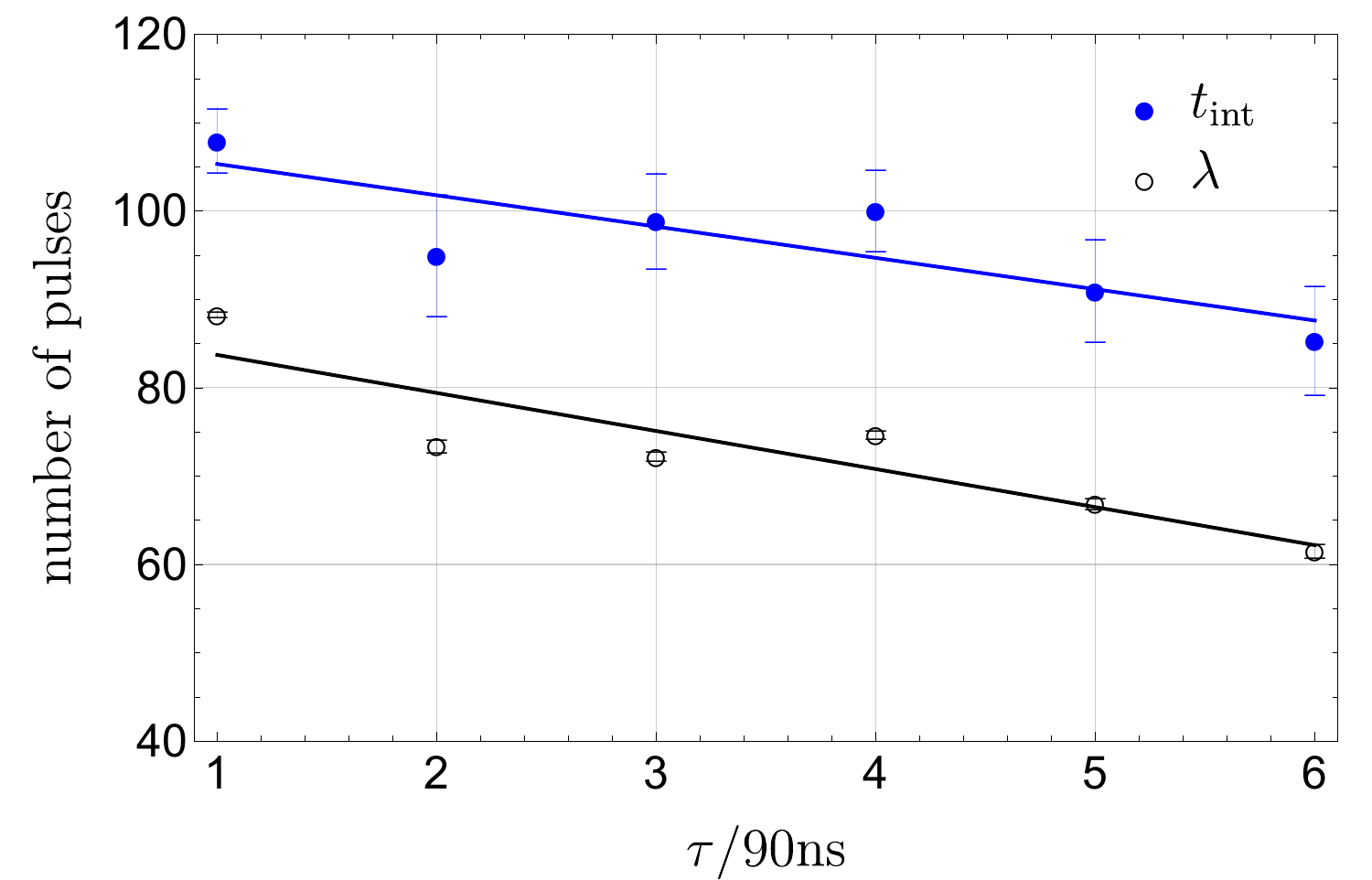}
\caption{(Color online) DD performance as a function of pulse interval $\tau$ for IBMQX5 (in units of $90$ns).
The intersection time $t_{\text{int}}$ of the fidelity curves under free evolution and DD, and the decay-time exponent $\lambda$, as a function of the pulse spacing, $\tau$. Linear fits yields $t_{\mathrm{int}} = -3.5 (\tau / 90 \text{ns}) + 108$ and $\lambda = -4.3 (\tau /90 \text{ns} ) + 88.0$. 
}
\label{fig:timeEvol}
\end{figure}

\textit{Dependence on the pulse interval.}---%
It is well known from DD theory that performance depends strongly on the pulse interval $\tau$~\cite{lidar2013quantum}. We thus consider the intersection time of the fidelity curves under free evolution and DD for IBMQX5, denoted $t_{\text{int}}$, which represents the duration over which DD improves the fidelity over free evolution. The dependence on $\tau$ is shown in Fig.~\ref{fig:timeEvol}. We observe that, as expected, $t_{\text{int}}$ decays to first order in $\tau$, implying that as the pulse interval increases, DD becomes less effective. Also shown in Fig.~\ref{fig:timeEvol} is the decay-time exponent $\lambda$ as a function of $\tau$, which behaves similarly: $\lambda$ decays from an initial value of $\approx 88$ to $\approx 60$, still twice as large as that of free evolution ($\approx 29$). Somewhat surprisingly, both $t_{\text{int}}$ and $\lambda$ decay non-monotonically with $\tau$, a finding that is not captured by standard DD theory and presents an interesting open theoretical problem that is beyond the scope of this work.

\begin{figure}[t]
	\raggedright
	\includegraphics[width = .48\textwidth]{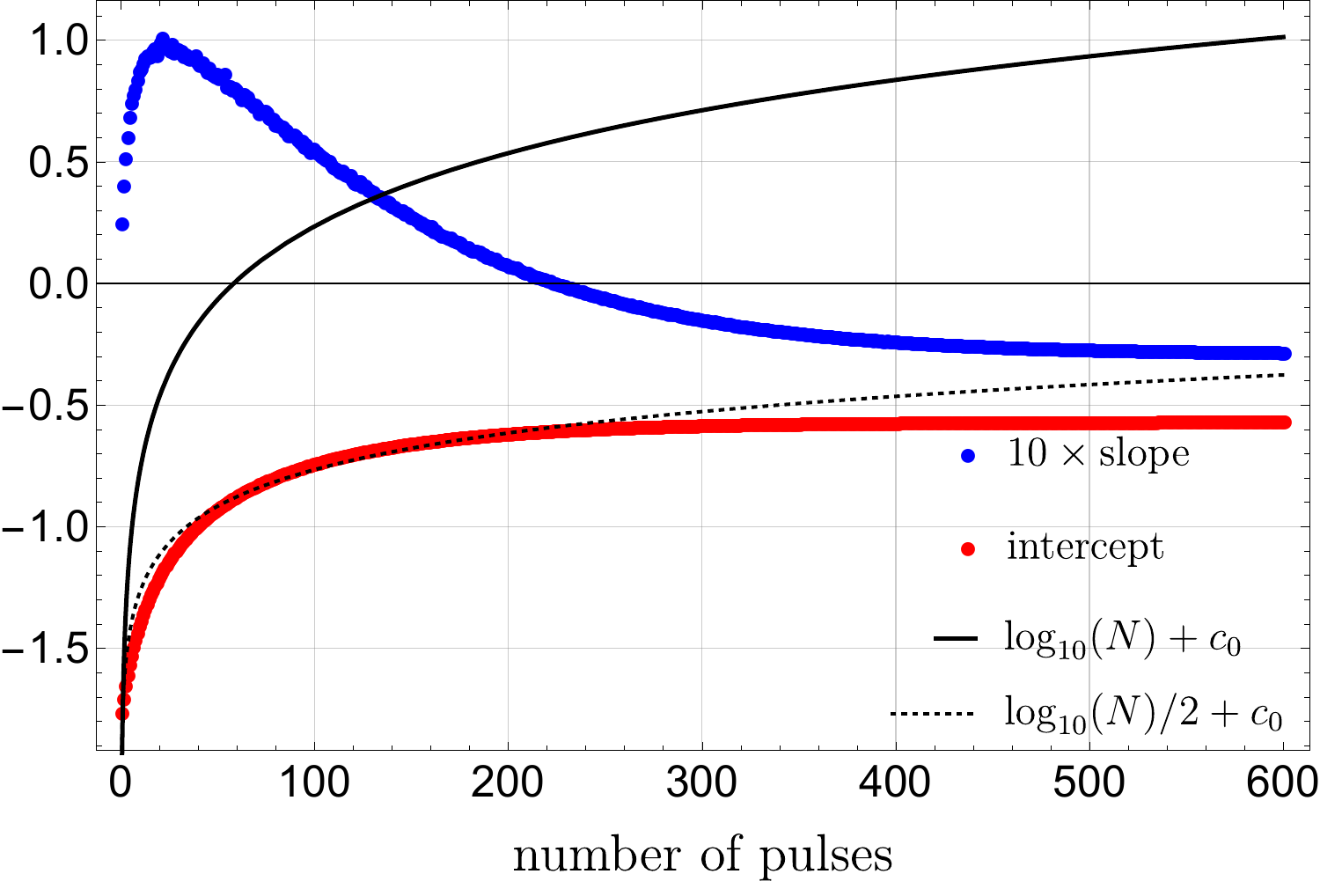}
	\caption[ ]{(Color online) The slope $a$ ($\times 10$) and intercept $b\log N$ derived from plotting $\log ( \sqrt{1 -F})$ as a function of $\log(\tau)$ for IBMQX5, at different numbers of pulses $N$ (see Appendix~\ref{app:dist} for more details). The solid black line is $\log(N) + c_0$ and the dotted black line is $\log(N)/2 + c_0$, where $c_0$ is the intercept at $N=0$.}
	\label{fig:distanceSlopeIntercept}
\end{figure}

For ideal DD pulses the infidelity between the free and dynamically decoupled states is bounded as~\cite{daniel2008distance,daniel2007performance} (see also Appendix~\ref{app:dist}):
\begin{equation}
 \log( 1 -\sqrt{F}) \leq a \log(\tau) + b \log(N) + c.
\label{eq:infid}
\end{equation}
where 
$c$ is a constant directly related to the operator norms of the bath and system-bath Hamiltonian, for ideal pulses $a=2$ and $b=1$, and 
$F$ is the fidelity~\cite{Uhlmann}: $F(\rho_1, \rho_2) = (\| \sqrt{\rho_1}\sqrt{\rho_2}\|_1)^2$, with $\rho_1(T)$ and $\rho_2(T)$ being the state of the system under DD and free evolution, after total evolution time $T=N\tau$. 
%
%
The bound~\eqref{eq:infid} is checked in Fig.~\ref{fig:distanceSlopeIntercept}. The slope $a$, capturing the pulse interval dependence, is significantly smaller than the theoretical upper bound of $2$ given in Eq.~\eqref{eq:infid}, but consistent with it. The offset $b\log N$, capturing the pulse number dependence, is also consistent with the theoretical upper bound with $b=1$ (in fact, it more closely matches $b=1/2$). It is not surprising that the bound is not tight in the realistic case of finite-width pulses, but it is interesting that the slope $a$ becomes negative for sufficiently large $N$ ($\gtrsim 200$). The interpretation of the bound~\eqref{eq:infid} for $a>0$ is that increasing $N$ or $\tau$ while keeping the other variable fixed increases the infidelity bound, i.e., is expected to reduce DD's performance. However, a negative slope in fact implies the opposite: a decreasing infidelity with increasing pulse interval $\tau$, at fixed $N$. This too is a nonstandard finding and presents another interesting open theoretical problem.

\begin{figure}[h]
\centering
\subfigure
{
 \Qcircuit @C=1em @R=0.7em @!R {
& \lstick{\ket{0}} & \qw & \gate{H} & \ctrl{1} &  \gate{X} & \gate{Y} & \gate{X} & \gate{Y}  & \measureD{Z} \\
& \lstick{\ket{0}} & \qw & \qw      & \targ    &  \gate{X} & \gate{Y} & \gate{X} & \gate{Y}  & \measureD{Z}
\label{circuit:BellState1}
}
}
\subfigure
{
\Qcircuit @C=1em @R=0.7em @!R {
& \lstick{\ket{0}} & \qw & \gate{H} & \ctrl{1} & \qw       &  \gate{X} & \gate{Y} & \gate{X} & \gate{Y}  & \measureD{Z} \\
& \lstick{\ket{0}} & \qw & \qw      & \targ    &  \gate{X} &  \gate{X} & \gate{Y} & \gate{X} & \gate{Y}  & \measureD{Z}
}    
\label{circuit:BellState1}
}
\caption[ ]{Quantum circuits used to initialize $\ket{\Phi^+}$ (top) and $\ket{\Psi^+}$ (bottom) on the IBMQX5, followed by a DD pulse sequence (only a single repetition is shown), and measurement in the computational basis. }
\label{figure:BellState}
\end{figure}
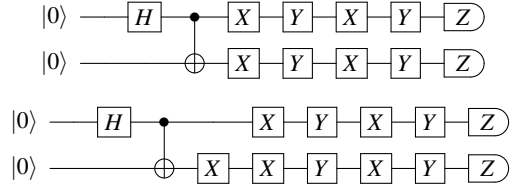

\textit{Protection of two-qubit entangled states.}---%
To evaluate the performance of DD in preserving entangled states, we initialized qubit pairs in Bell states of the form
$\ket{\Phi^+} = \frac{1}{\sqrt{2}} (| 00 \rangle + \ket{11})$ and $\ket{\Psi^+} = \frac{1}{\sqrt{2}} (\ket{01} + \ket{10})$,
followed by an XY4 DD sequence (higher order DD sequences for entanglement protection are known as well~\cite{Mukhtar:10}). 
Ideally, one would perform the measurements in the Bell basis and report the corresponding fidelities. 
However, we found that due to the relatively large errors introduced by CNOT gates and the high readout errors, Bell basis measurements yielded very noisy data which was difficult to draw meaningful conclusions from. Therefore we instead performed a measurement of both qubits in the computational basis $\{| 00 \rangle, \ket{01}, \ket{10}, |11 \rangle\}$, as illustrated in 
Fig.~\ref{figure:BellState}.  

\begin{figure*}[ht]

\hspace{-6mm}
\includegraphics[scale=0.36]{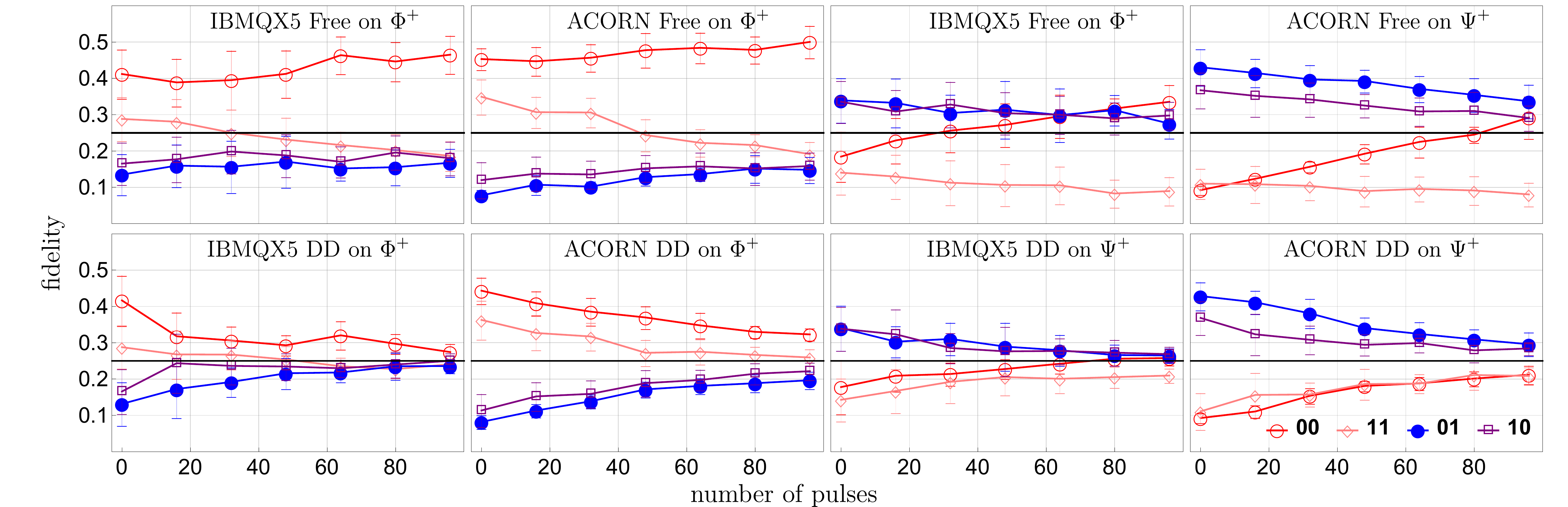}
\caption[ ]{(Color online)  Probabilities of different computational basis states after DD for initially prepared Bell states $\ket{\Phi^+} = \frac{1}{\sqrt{2}} (| 00 \rangle + \ket{11})$ and $\ket{\Psi^+} = \frac{1}{\sqrt{2}} (\ket{01} + \ket{10})$, as a function of the number of pulses, for both IBMQX5 and Acorn. 
Top row: free evolution. Bottom row: evolution under DD. The $\ket{00}$ state is favored under free evolution. The solid horizontal line indicates $p=0.25$, the limit of a fully mixed state. For $\ket{\Psi^+}$ there is no noticeable difference in the performance with or without DD. For $\ket{\Phi^+}$ on IBMQX5, after $N\approx 20$, $p_{11} \approx 0.25$, suggesting that at this point all information has essentially been scrambled. On Acorn, complete scrambling of $\ket{\Phi^+}$ occurs after $N \sim 30$. Overall, DD is more effective at slowing down the decay to the fully mixed state for Acorn than for IBMQX5.}
\label{fig:timeEvolBell}
\end{figure*}

Let $p_{ij}$ be the probability of measuring the computational basis state $\ket{ij}$, with $i,j\in\{0,1\}$. Our results are plotted in Fig.~\ref{fig:timeEvolBell}, which shows the probabilities $p_{ij}$ that were measured after initializing the system in a Bell state and letting it evolve either freely or under DD. Under ideal conditions one would expect to have $p_{00}=p_{11}=0.5$ for $\ket{\Phi^+}$ and $p_{01}=p_{10}=0.5$ for $\ket{\Psi^+}$. Instead, for both QCs, Fig.~\ref{fig:timeEvolBell} (top row) shows a strong bias for $\ket{00}$ over $\ket{11}$ upon initialization ($N=0$) for the $\ket{\Phi^+}$ case, with some contamination by the $\ket{01}$ and $\ket{10}$ states. For the $\ket{\Psi^+}$ case, Fig.~\ref{fig:timeEvolBell} (bottom row) shows contamination by $\ket{00}$ and $\ket{11}$ upon initialization (stronger for IBMQX5 than for Acorn), and a curious bias towards $\ket{01}$ over $\ket{10}$ for Acorn. We attribute these effects to the single-digit percentage readout errors (see Appendix~\ref{sub:machinespecs}) and the CNOT gate errors. Clearly, the preparation of the Bell states is itself prone to substantial errors on both QCs. 

As mentioned earlier, SE plays a key role and consequently the main effect under free evolution is a sharp increase in $p_{00}$ with $N$ on both devices. Under DD, the main beneficial effect is that this dominance of the ground state $\ket{00}$ is suppressed. However, on IBMQX5 for both $\ket{\Phi^+}$ and $\ket{\Psi^+}$ a nearly uniform distribution over all four computational basis states is reached after $100$ pulses. The trend is similar for Acorn, but the decay to the fully mixed state is slowed down more by DD than for IBMQX5, and DD manages to keep the original ratio of $\frac{p_{00}}{p_{11}}$ up to $\sim 50$ pulses. Overall, it is clear that entanglement is rapidly lost, but is slowed down somewhat by DD.

\textit{Conclusions and outlook.}---%
The results reported here demonstrate the undeniable usefulness of DD on prototype QCs for the suppression of inherent decoherence, a feature which has yet to be demonstrated unconditionally using QEC~\cite{ibm2016simon,2017repetition,ibm2017useful,2017cpc,2017simpleqec,2018arXivQFTsmall,Harper:2018aa} (see Appendix~\ref{app:QEC-review}). It is remarkable that performance improvement was achievable despite significant pulse implementation imperfections. Therefore, we conclude that, given a quantum circuit, it is already advantageous to perform dynamically decoupled evolution rather than free evolution between computational gates~\cite{Ng:2011dn}.


In the future, as the error rates of measurement and multi-qubit gates are reduced, it should become possible to more accurately assess the effectiveness of DD. We anticipate that reduction in multi-qubit errors will alleviate the restrictions placed by connectivity of the qubits as it will be possible to perform more SWAP gates without corrupting the states. In such scenarios, hybrid QEC-DD~\cite{KhodjastehLidar:03,Ng:2011dn,Paz-Silva:2013tt} methods could be experimentally assessed and would constitute an attractive near-term target for higher performance gains than is enabled by either scheme alone.

Another attractive prospect for future experiments is the implementation of higher-order DD sequences. Indeed, we have already tested higher-order sequences based on genetic algorithms~\cite{genetic}, and found a small improvement over XY4 (see Appendix~\ref{app:GA}). 
The success of such sequences in providing better fidelity improvements than the XY4 sequence will depend on improved pulse control (such as the ability to fine-tune pulse intervals, needed to implement UDD~\cite{udd} and QDD~\cite{qdd}), reduction of the pulse interval and duration, etc. Implementation of robust DD sequences~\cite{Souza:2011aa,Souza:2012aa,genetic,Kabytayev:2014aa,Genov:2017aa} is another particularly promising venue.

\textit{Acknowledgements.}---
We acknowledge the use of the IBM Quantum Experience and Rigetti's Forest for this work. The views expressed are those of the authors and do not reflect the official policy or position of IBM, the IBM Quantum Experience team, Rigetti, or the Rigetti team. This work was supported in part by Oracle Labs, part of Oracle America, Inc. which provided the funding for the donation in support of this academic research. We are grateful to Haimeng Zhang for insightful discussions. 

\bibliographystyle{apsrev4-1}
\bibliography{ibm,refs}

\newpage

\appendix

\section{Overview of existing QEC implementations on the IBM Quantum Experience}
\label{app:QEC-review}

Several quantum error correcting and detecting techniques have been implemented on the IBM Quantum Experience (IBM-QE) chips, including both the QX4 ($5$ qubits) and QX5 ($16$ qubits), albeit with limited success. 

Devitt attempted to implement an error-corrected Rabi oscillation across a logically encoded qubit using a distance-two surface code \cite{ibm2016simon}. However, since it was not possible to prepare a logical state of the form $e^{i \theta X_{L}}|0_{L}\rangle$, he prepared a single qubit in the rotated state and then encoded it (which is not a fault-tolerant operation). The encoded but {uncorrected} qubit showed almost no Rabi oscillation while the corrected qubit displayed a very low-visibility oscillation. The error-corrected version did not outperform a non-error-corrected version of the same experiment.   

Wootton and Loss demonstrated a repetition code of $15$ qubits \cite{2017repetition} and showed that the logical error rate decays exponentially with code distance. This is a classical repetition code that can detect and correct only one error type, i.e., it corrects a logical \emph{bit} instead of a logical qubit.

Sohn \textit{et. al} implemented an encoded memory qubit \cite{2017simpleqec} -- using a simple $3$-qubit code. They considered {artificial} bit-flip and phase-flip errors, and used idle gates to prolong the memory time. The encoded qubit fidelities reported were lower than those of the physical qubits.

The syndrome information produced by a $[[4,2,2]]$ coherent parity check (CPC) quantum memory can be used to improve the fidelity and purity of the code output. Using post-selection, Roffe \textit{et al}. \cite{2017cpc} showed that the fidelity increased from $0.62 \pm 0.03$ to $0.75 \pm 0.04$ (with $1 \sigma$ error bars) but with an average yield of $(54 \pm 2 \%)$ (the average yield is the proportion of data retained after post-selection, averaged over several runs).

Sampling from $20$ different states that the $[[4,2,2]]$ code can fault-tolerantly prepare was demonstrated by Vuillot~\cite{ibm2017useful}. This is based on a recent proposal by Gottesman who defined a new criterion for experimental fault-tolerance that can be demonstrated using $5$ qubits arranged on a ring with nearest-neighbor interactions~\cite{gottesman2017small}. Essentially, the criterion is: if the error rate for an encoded circuit is less than that of the unencoded circuit, \emph{for all circuits} in the family of circuits of interest, then this counts as a valid demonstration of fault-tolerance for a small quantum system. Since the $[[4,2,2]]$ code is error-detecting but not correcting, it can only improve error rates via post-selection. Since post-selection is common for ancilla preparation subroutines in large fault-tolerant protocols, a successful demonstration of fault-tolerance with the $[[4,2,2]]$ code can be considered a demonstration of \emph{fault-tolerant ancilla preparation}. In Vuillot's demonstration, there were two fault-tolerant versions, ``FTv1'' and ``FTv2'', where FTv1 has the lowest average error, but with a post-selection ratio of $0.65$ (that is, data is kept only for 65\% of the trials) and FTv2 has the worst performance (in terms of average error) with a post-selection ratio of $0.44$. This result was ambiguous as to the efficacy of the circuits for fault tolerance, due to the lack of an appropriate success metric. The results also depended on initial states, circuit length, and type. 
 
Another implementation of fault-tolerance for a class of circuits with post-selection ratios around $\approx 0.6$ was performed by Willsch \textit{et al}.~\cite{2018arXivQFTsmall}. The authors note that the fault-tolerance criterion was not satisfied on all days that they ran the experiment and since the errors present in an actual application can be much more complicated than those assumed in the design of the protocol, it was not guaranteed that using a fault-tolerant protocol improves the computation.

Harper and Flammia \cite{Harper:2018aa} attempted to demonstrate the error detection capability of  the $[[4,2,2]]$ code described by Gottesman~\cite{gottesman2017small} on the IBMQX5 by measuring the average gate fidelity ($F$) for this code using randomized benchmarking (RB) over a \emph{subset} of the Clifford group. Their measurements yielded an infidelity ($1-F$) of $5.8 \%$ without any encoding versus $0.6\%$ infidelity for the encoded case when using a selected subset of IBMQX5 qubits, with a post-selection ratio of $\approx 40 \%$. The infidelity for the unencoded $2$-qubit case was obtained by applying gates from a gate set ``GS1'' (in random circuits) to the physical ground state $| 00 \rangle$ and a ``phased" state which is obtained by the application of Hadamard gates to both qubits followed by the phase gate on the first qubit of the $| 00 \rangle$ state. For the encoded $4$-qubit case, there are two initial states: the physical ground state $| 0000 \rangle$ which substitutes for the logical ground state, and an encoded ``phased'' state, which is prepared from $| 0000 \rangle$ instead of $| \overline{00} \rangle$. The gate set ``GS2'' (which is logically equivalent to GS1) was then applied in random circuits. After applying the gates from the corresponding gate sets, an inversion gate was applied to return each qubit to its initial state, and the fidelity of the final state with respect to the initial state was measured. Finally, all solutions that have odd parity were discarded, as the $[[4,2,2]]$ code is supported only over even parity states. The post-selected fidelities from the physical ground state and the phased state were then used to calculate the average gate infidelity.


We offer four reasons to be critical of the results in Ref.~\cite{Harper:2018aa}. (i) An ideal RB experiment would first prepare the logical ground state (or some other logical state) of the codespace. However, citing the robustness of RB to the high error rates for state preparation and measurement (SPAM) errors, the logical state preparation was skipped entirely.  Note that neither the physical ground state nor the ``phased" state, on which RB was performed, are in the codespace of the $[[4,2,2]]$ code, whose stabilizers are $XXXX$ and $ZZZZ$. (ii) Recall that the $|0000 \rangle$ state is inherently robust since it does not spontaneously emit. Preparation of other logical states involving physical qubits in the (excited) $\ket{1}$ state was not attempted, and therefore it is difficult to generalize from these results. We believe that showing robustness to SPAM errors by calculating the infidelity after preparing \emph{different} logical states would have made the results more compelling. (iii) The gate sets were cleverly chosen such that GS2 only includes local single qubit operations like $XXXX$ and SWAPs (which can be performed virtually, as opposed to physically, thus avoiding gate-induced noise by simply relabelling the qubits), while GS1 comprises local gates and also entangling gates, like CNOT and CZ gates. As a result, the gate sets favor the encoded case while handicapping the unencoded one. The authors do note that more than half of the change in infidelity is because the encoded case can altogether \emph{bypass} multi-qubit gates, which are substantially noisier compared to single qubit gates. The remaining change in infidelity (about $2 \%$) comes from the ability to perform post-selection. (iv) Proctor \textit{et. al.} \cite{rb2017} have recently argued that although the average gate infidelity obtained by RB, $r$, has some physical relevance, it does not correspond to the actual \emph{experimental} gate infidelity, $\epsilon$. Their results also show that $r$ can be both greater or smaller than $\epsilon$, implying that it is neither an upper nor a lower bound on $\epsilon$, which, in turn, affects the ability of RB to characterize noise in  QCs. While the results of Ref.~\cite{Harper:2018aa} provide an incentive to further investigate error detection in near-term devices, a conclusive demonstration of the advantage of codes such as the $[[4,2,2]]$ code will need to overcome the critique raised above.

\section{Machine Specifications}
\label{sub:machinespecs}

The IBMQX5 and Rigetti Acorn chips used in our experiments have similar figures of merit, including $\textit{T}_1$ and $\textit{T}_2$ times, single-qubit gate errors, readout errors, etc. Details and additional information regarding the physical parameters of the systems are provided in Tables~\ref{tab:IBMQX5-Paramters} and \ref{tab:Acorn}. Since these parameters fluctuate on a daily basis a date of access is also included. Yet more information is summarized in Table~\ref{table:params}.

\begin{table}[!]
\centering
\resizebox{\columnwidth}{!}{
\begin{tabular}{|c|c|c|c|c|c|c|}
\hline
Qubit & $T_1$ {[}$\mu$s{]} & $T_2$ {[}$\mu$s{]} & \begin{tabular}[c]{@{}c@{}}Gate error\\   {[}10\textasciicircum -3{]}\end{tabular} & \begin{tabular}[c]{@{}c@{}}Readout Error\\   {[}10\textasciicircum -2{]}\end{tabular} & Gate Fidelity & Readout Fidelity \\ \hline
0 & 47.0 & 29.4 & 1.95 & 4.78 & 0.9980 & 0.9522 \\ \hline
1 & 35.1 & 54.7 & 3.82 & 5.00 & 0.9962 & 0.9500 \\ \hline
2 & 35.7 & 43.3 & 3.72 & 4.45 & 0.9963 & 0.9554 \\ \hline
3 & 54.3 & 80.3 & 2.25 & 8.95 & 0.9977 & 0.9105 \\ \hline
4 & 39.3 & 44.7 & 2.13 & 7.76 & 0.9979 & 0.9224 \\ \hline
5 & 43.3 & 57.2 & 1.68 & 5.83 & 0.9983 & 0.9417 \\ \hline
6 & 55.2 & 91.7 & 2.37 & 4.10 & 0.9976 & 0.9589 \\ \hline
7 & 28.9 & 27.8 & 3.17 & 4.01 & 0.9968 & 0.9599 \\ \hline
8 & 59.5 & 101.6 & 1.13 & 5.86 & 0.9989 & 0.9413 \\ \hline
9 & 48.6 & 82.9 & 1.10 & 11.37 & 0.9989 & 0.8862 \\ \hline
10 & 27.5 & 40.9 & 4.41 & 11.76 & 0.9956 & 0.8824 \\ \hline
11 & 57.3 & 102 & 1.81 & 5.03 & 0.9982 & 0.9497 \\ \hline
12 & 47.5 & 55 & 1.39 & 13.24 & 0.9986 & 0.8676 \\ \hline
13 & 51.8 & 97.1 & 1.63 & 4.25 & 0.9984 & 0.9574 \\ \hline
14 & 40.6 & 72.3 & 2.11 & 6.51 & 0.9979 & 0.9349 \\ \hline
15 & 37.3 & 72.8 & 3.90 & 10.53 & 0.9961 & 0.8946 \\ \hline
Mean & 44.3 & 70.0 & 2.41 & 7.09 & 0.9976 & 0.9291 \\ \hline
SD & 7.4 & 19.1 & 1.06 & 3.10 & 0.0011 & 0.0310 \\ \hline
\end{tabular}
}
\caption{Physical Parameters-IBMQX5 - Accessed 06/19/2018. The minimum, average, and maximum CNOT gate fidelity are 0.8417, 0.9330, and 0.9513 respectively. The gate (readout) fidelity is $1$ minus gate (readout) error.}
\label{tab:IBMQX5-Paramters}
\end{table}

\begin{table}[!]
\centering
\resizebox{\columnwidth}{!}{
\begin{tabular}{|c|c|c|c|c|c|c|}
	
	\hline
	Qubit & $T_1$ {[}$\mu$s{]} & $T_2^*$ {[}$\mu$s{]} & \begin{tabular}[c]{@{}c@{}}Gate error\\   {[}10\textasciicircum -3{]}\end{tabular} & \begin{tabular}[c]{@{}c@{}}Readout Error\\   {[}10\textasciicircum -2{]}\end{tabular} & Gate Fidelity & Readout Fidelity \\ \hline
	0 & 15.2 & 7.2 &  0.1 &  4.99 & 0.9999 & 0.9501 \\ \hline
	1 & 17.6 & 7.7 &  0.1 &  4.86 & 0.9999 & 0.9514 \\ \hline
	2 & 18.2 & 10.8 & 4.3 &  25.54 & 0.9957 & 0.7446 \\ \hline
	3 & 31.0 & 16.8 & 9.2 &  11.4 & 0.9908 & 0.886 \\ \hline
	4 & 23.0 & 5.2 & 16.0 &  15.35 & 0.984 & 0.8465 \\ \hline
	5 & 22.2 & 11.1 &  14.1 &  15.57 & 0.9859 & 0.8443 \\ \hline
	6 & 26.8 & 26.8 & 25.1 & 17.83 & 0.9749 & 0.8217 \\ \hline
	7 & 29.4 & 13.0 &  14.1 &  7.08 & 0.9859 & 0.9292 \\ \hline
	8 & 24.5 & 13.8 &  15.6 &  4.74 & 0.9844 & 0.9526 \\ \hline
	9 & 20.8 & 13.8 &  42.3 &  11.3 & 0.9577 & 0.887 \\ \hline
	10 & 17.1 & 10.6 &  15.0 &  5.18 & 0.985 & 0.9482 \\ \hline
	11 & 16.9 & 4.9 &  27.0 & 2.7 & 0.973 & 0.973 \\ \hline
	12 & 8.2 & 10.9 &  13.0 &  3.75 & 0.987 & 0.9625 \\ \hline
	13 & 18.7 & 12.7 &  28.0 & 3.65 & 0.972 & 0.9635 \\ \hline
	14 & 13.9 & 9.4 &  16.0 &  4.43 & 0.984 & 0.9557 \\ \hline
	15 & 20.8 & 7.3 & 18.0 &  19.68 & 0.982 & 0.8032 \\ \hline
	16 & 16.7 & 7.5 &  30.0 &  5.87 & 0.97 & 0.9413 \\ \hline
	17 & 24.0 & 8.4 &  21.4 &  4.02 & 0.9786 & 0.9598 \\ \hline
	18 & 16.9 & 12.9 &  33.7 &  6.95 & 0.9598 & 0.9305 \\ \hline
	19 & 24.7 & 9.8 &  13.7 &  5.12 & 0.9863 & 0.9488 \\ \hline
	Mean & 20.33 & 11.03 & 17.83 &  9.001 & 0.9822 & 0.9065 \\ \hline
	SD & 5.50 & 4.83 & 10.82 &  6.49 & 0.0108 & 0.0632 \\ \hline
	
\end{tabular}
}
\caption{Physical parameters for Rigetti Acorn - Accessed 4/13/18.  The minimum, average, and maximum Controlled-Z gate fidelity are 0.72, 0.865, and 0.917 respectively. The gate (readout) fidelity is $1$ minus gate (readout) error.}
\label{tab:Acorn}
\end{table}

\begin{table}
\centering
\resizebox{\columnwidth}{!}{
 \begin{tabular}{||c c c||} 
 \hline
 Parameter & IBMQX5 (IBM) & 19Q-Acorn (Rigetti) \\ [0.5ex] 
 \hline\hline
 Single qubit pulse time (in ns) & 90 & 100 \\
 Number of qubits & 16 & 19 (15 used) \\
 Shots per experiment & 8192 & 1000 \\
 Interface & QASM & pyQuil \\
 [1ex] 
 \hline
\end{tabular}
}
\caption{Single qubit pulse times, number of qubits, and shots per experiment for the IBMQX5 and 19Q-Acorn.}
\label{table:params}
\end{table}

\subsection{IBMQX5}

\begin{figure}[t]
\centering
\includegraphics[scale=0.37]{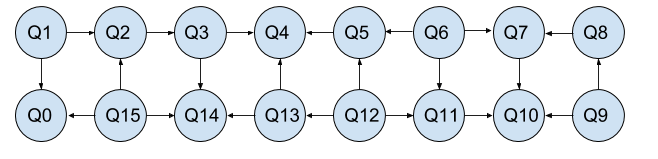}
    \label{fig:topologyx5}
\caption[ ]{(Color online) Connectivity between the qubits in IBMQX5. 
The qubit at the tail (tip) of an arrow is the control (target) in a controlled-$U$ gate.}
\label{fig:topology}
\end{figure}

We ran experiments on this $16$-qubit chip with circuits written in Open Quantum Assembly Language \cite{cross2017open}. The qubit connectivity graph is illustrated in Fig.~\ref{fig:topology}, with the corresponding cross-talk and relaxation times reported in \cite{IBMQX5}. The chip comprises superconducting transmon qubits, coupled via coplanar waveguides. Each single-qubit pulse took $90$ns to implement (including the buffer time of $10$ns), with the states evolving freely between the pulses, and each experiment was performed $8192$ times. The single qubit gate error and readout errors are on the order of $10^{-3}$.
Dephasing times ($T_2)$ for the qubits were between $30 - 100 \mu$s with a set of times given in Table \ref{tab:IBMQX5-Paramters}. The results reported here were from experiments done during the period 6/19/18-6/22/18 unless stated otherwise. For the Bell-state experiments in Figure \ref{fig:timeEvolBell}, the following qubits were paired together: $(0,1), (2, 15), (3, 14), (5, 12), (6, 11), (7, 10), (8,9)$ and the experiments were performed on 3/14/18. Substantial daily variations are common, but our results are qualitatively robust to such variations.

\begin{table}[!]
\centering
\resizebox{\columnwidth}{!}{
\begin{tabular}{|c|c|c|c|c|c|c|}
\hline
Qubit & $T_1$ {[}$\mu$s{]} & $T_2$ {[}$\mu$s{]} & \begin{tabular}[c]{@{}c@{}}Gate error\\   {[}10\textasciicircum -3{]}\end{tabular} & \begin{tabular}[c]{@{}c@{}}Readout Error\\   {[}10\textasciicircum -2{]}\end{tabular} & Gate Fidelity & Readout Fidelity \\ \hline
0 & 50.8 & 14.7 & 0.86 & 4.80 & 0.9991 & 0.9520 \\ \hline
1 & 50.0 & 64.6 & 1.46 & 5.30 & 0.9985 & 0.9470 \\ \hline
2 & 47.9 & 45.0 & 1.29 & 9.80 & 0.9987 & 0.9020 \\ \hline
3 & 37.4 & 15.1 & 3.44 & 5.70 & 0.9966 & 0.9430 \\ \hline
4 & 56.0 & 30.5 & 0.94 & 7.00 & 0.9991 & 0.9300 \\ \hline
Mean & 48.4 & 34.0 & 1.60 & 6.52 & 0.9984 & 0.9348 \\ \hline
SD & 6.8 & 21.2 & 1.06 & 2.01 & 0.0010 & 0.0201 \\ \hline
\end{tabular}
}
\caption{Physical Parameters for IBMQX4 - Accessed 06/21/2018. The minimum, average, and maximum CNOT gate fidelity are 0.8738, 0.9441, and 0.9774 respectively. The gate (readout) fidelity is $1$ minus gate (readout) error.}
\label{tab:IBMQX4-Parameters}
\end{table}

\subsection{IBMQX4}
\label{sec:app:QX4}
\begin{figure}[t]
\centering
\includegraphics[scale=0.37]{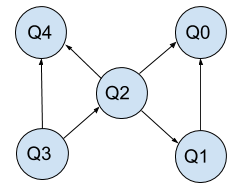}
\caption[ ]{(Color online) Connectivity between the qubits in IBMQX4. 
The qubit at the tail (tip) of an arrow is the control (target) in a controlled-$U$ gate.}
\label{fig:topology-qx4}
\end{figure}

We also ran experiments on the $5$-qubit IBMQX4 chip, whose connectivity graph is shown in Fig.~\ref{fig:topology-qx4}. Each single-qubit pulse took $60$ns to implement (including the buffer time of $10$ns), with the states evolving freely between the pulses, and each experiment was performed $8192$ times. The single qubit gate error and readout errors are on the order of $10^{-3}$. Dephasing times ($T_2)$ for the qubits were between $30 - 100 \mu$s with a set of times given in Table \ref{tab:IBMQX4-Parameters}. The results reported here were from experiments done during the period 6/19/18-6/21/18.

\begin{figure}[t]
	\raggedright
	\includegraphics[scale=0.6]{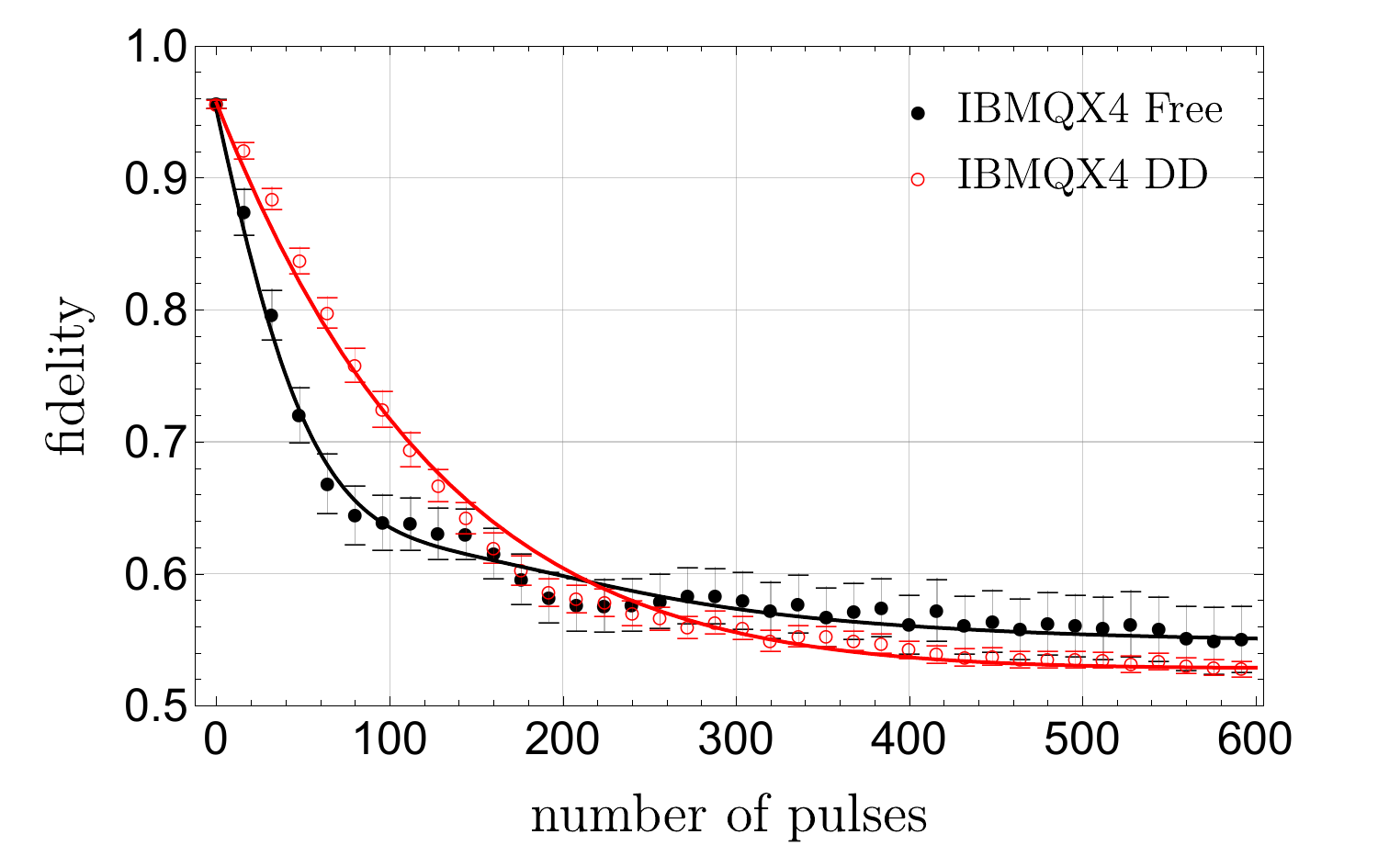}
	\caption[ ]{(Color online) IBMQX4 results for the fidelities under DD compared to free evolution, after averaging over $36$ initial conditions (type 2) and all $5$ qubits. }
	\label{fig:x4dd}
\end{figure}

We tested the performance of the XY4 sequence on the IBMQX4. As shown in Table~\ref{tab:resultsx4rigetti} and Fig.~\ref{fig:x4dd}, we found improvements that are qualitatively consistent with the IBMQX5 results. We again see both a dominance of Markovian exponential decay along with a nearly $3$-fold increase in $\lambda$, from $44.7 \pm 2.8$ under free evolution to $128.0 \pm 0.8$ under DD. The intersection between free and DD evolution occurs at $t_{\mathrm{int}} = 216 \pm 16$, which is twice the gate depth for which DD improves the average fidelity on IBMQX5.

\subsection{Rigetti Acorn}
\begin{figure}[t]
\centering
{
   \includegraphics[scale=0.25]{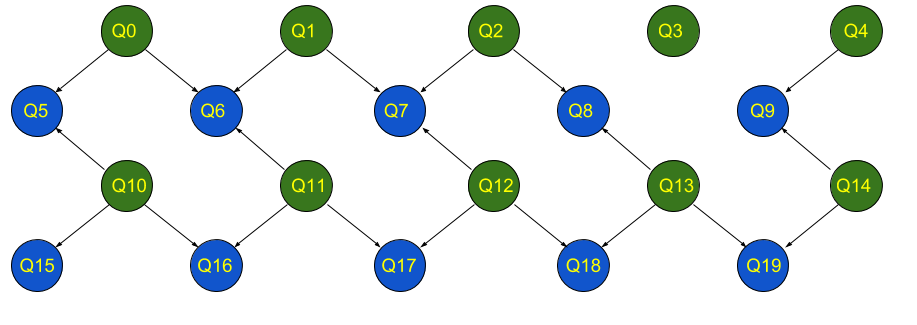}
}

\caption[ ]{(Color online) Connectivity between the qubits within Acorn. Qubit 3 is disconnected. Qubits 2,12, 15,18 were not used as their performance varied substantially over time. Based on \cite{RigettiSpecs}.}
\label{fig:acornTopology}
\end{figure}

The $19$-qubit Acorn chip was accessed via Rigetti's Forest, using circuits written in Quil \cite{RigettiQuil} with remote access provided through the pyQuil interface \cite{RigettiSpecs}. Acorn comprises a combination of fixed frequency (qubits 0-4 and 10-14) and tunable (qubits 5-9 and 15-19) transmon qubits, capacitive coupled. Within the current layout, qubit 3 was disabled due to performance issues, leaving $19$ qubits functional for programming. The single qubit gate time was $50$ns (for all qubits other than 2, 18 which have gate times $100$ ns) with dephasing times ($T_2^*)$ varying between $10 - 40 \mu$s as summarized in Table \ref{tab:Acorn}. The connectivity graph is illustrated in Fig.~\ref{fig:acornTopology}.

Readout error was on the level of the IBM chip; while the relaxation times presented for the IBMQX5 were longer than those for Acorn, both chips exhibited relaxation times within the microsecond range with single-qubit fidelities greater than $0.98$.

\begin{figure}[t]
\raggedright
\includegraphics[width = .5\textwidth]{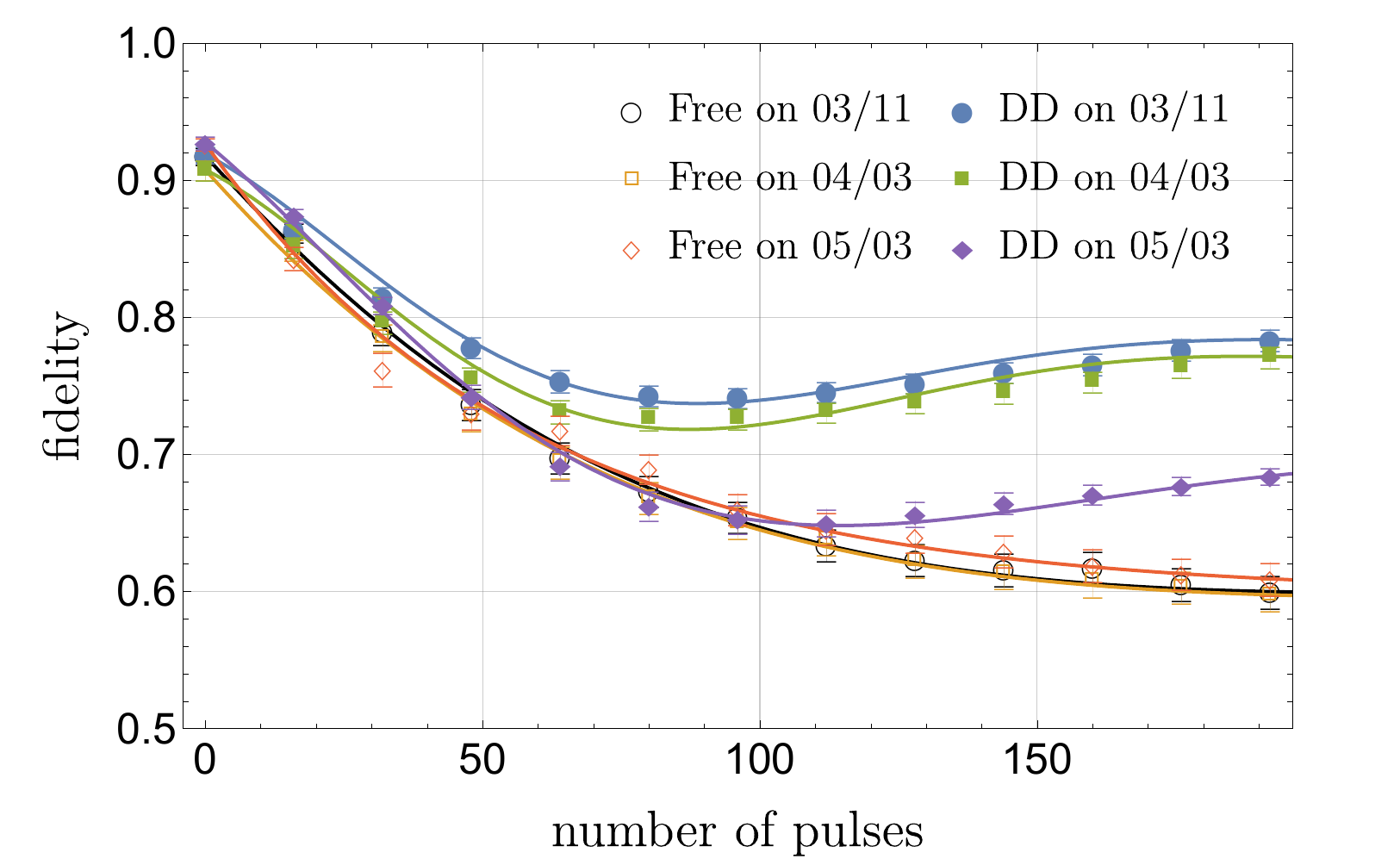}
\caption[ ]{(Color online) Acorn chip results for the mean fidelity averages over $36$ initial conditions (type 2) and all $15$ active qubits for different dates.} 
  \label{fig:timeEvolRigetti}
\end{figure}

We accessed the Acorn chip multiple times between 3/1/18 and 5/9/18. The results  reported here are from 4/3/18 unless stated otherwise. For the Bell-state experiments in Figure \ref{fig:timeEvolBell}, the following qubits were paired together: $(0,5), (1, 6), (4, 9), (11, 16), (12, 17), (14, 19)$ and the experiments were performed on 3/21/18.
Acorn's cryogenic system housing suffered a failure (a broken scroll pump) and the chip was out of service from 4/17/18 onwards \cite{acorn-warmup}. During the repair, the chip was brought back to room temperature and this thermal cycle and other possible contamination significantly affected performance, making qubits 2,3,15,18 unusable and also affecting multiple two-qubit gates (qubit 12 was very noisy and we did not use it). Owing to these factors, we only used $15$ total qubits in our analyses. In Fig.~\ref{fig:timeEvolRigetti} and Table~\ref{tab:resultsx4rigetti} we show how performance was affected before and after this repair. The main effect was a decrease in the final fidelity, due to a smaller recurrence, as captured by the smaller $\gamma$ value at the later date.

\begin{table}
\centering
\resizebox{\columnwidth}{!}{
 \begin{tabular}{|| c c c c c c c c ||}
 \hline
Machine & Accessed & Evolution & $F_0 \times 10^{-2}$ &$F_{N_{max}}  \times 10^{-2}$ & $\lambda$ & $\alpha$ &$\gamma$  \\
\hline\hline

Acorn  & 03/11/18 & Free & $91.8 \pm 0.3$ & $60.0 \pm  0.6$ & $69.7 \pm 1.3$ & $\infty$    & $0.16 \pm 0.09$ \\
Acorn  & 04/03/18 & Free & $90.8 \pm 0.4$ & $59.8 \pm  0.6$ & $68.1 \pm 1.3$ & $\infty$    & $0.14 \pm 0.11$ \\
Acorn  & 05/03/18 & Free & $92.7 \pm 0.2$ & $60.9 \pm  0.6$ & $57.2 \pm 1.2$ & $\infty$    & $0$             \\
Acorn  & 03/11/18 & DD   & $91.9 \pm 0.3$ & $78.4 \pm  0.4$ & $71.3 \pm 0.9$ & $\infty$    & $0.49 \pm 0.03 $ \\
Acorn  & 04/03/18 & DD   & $90.8 \pm 0.4$ & $77.1 \pm  0.4$ & $74.9 \pm 0.9$ & $\infty$    & $0.50 \pm 0.03 $ \\
Acorn  & 05/03/18 & DD   & $92.7 \pm 0.2$ & $68.4 \pm  0.3$ & $72.6 \pm 1.1$ & $\infty$    & $0.36 \pm 0.02 $ \\
IBMQX4 & 06/21/18 & Free & $95.7 \pm 0.2$ & $55.1 \pm  0.3$ & $44.7 \pm 2.8$ & $145 \pm 2$ & $0.37 \pm 0.13$ \\
IBMQX4 & 06/21/18 & DD   & $95.7 \pm 0.2$ & $52.9 \pm  0.3$ & $128. \pm 0.8$ & $\infty$    & $0.05 \pm 0.02$ \\

\hline \hline
\end{tabular}
}
\caption{Fit parameters for Acorn (considering only the $15$ active qubits) and IBMQX4 when Eq.~\eqref{eq:fit} is used to fit the mean fidelities in Fig.~\ref{fig:timeEvolRigetti}. }
\label{tab:resultsx4rigetti}
\end{table}

\section{Statistical methods}
\label{sub:boot}

\begin{figure}[t]
\centering
\hspace{-0.5cm}
\includegraphics[width = .48\textwidth]{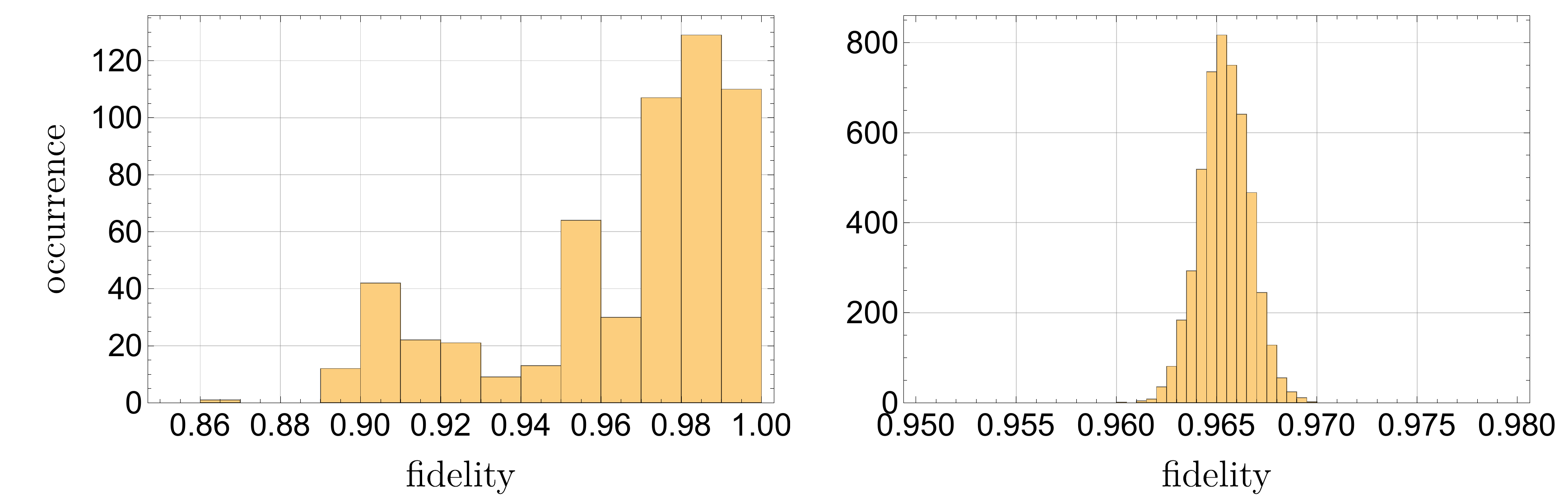}
\caption{(Color online)  Bootstrapping example: The left plot represents frequency counts (y-axis) versus fidelity (x-axis) for  a data set from the IBMQX5, taken from $36$ different initial conditions (type 1) and $16$ qubits. The right plot represents the samples after bootstrapping the original data set. The mean and confidence intervals were then calculated based on the bootstrapped distribution (right plot).}
\label{fig:bootstrap}
\end{figure}

To speed up data collection our DD experiments were performed in parallel on all qubits (or qubit pairs for the entanglement experiments) in the respective QCs. The mean values and error bars reported were computed after averaging over all qubits and subsequent bootstrapping.

The general bootstrapping technique implemented was based on Ref.~\cite{efron1992bootstrap}. Bootstrapping a data set was performed by taking the mean of $N$ resamples of $x$ points from the data set (with replacement). This generated a second representative set of data ($N$ large) from which the mean, standard deviation, and confidence intervals were then calculated. An example is given in Fig.~\ref{fig:bootstrap}. For the bootstrapped data presented here, the data was resampled $5000$ times (with each sample being the same size as the original data) with $576$ ($36$ initial conditions $\times 16$ qubits) samples drawn from the IBMQX5 data. 

In Fig.~\ref{fig:timeEvol} for the intersection time, $t_{\text{int}}$, the fidelity curves, $F(N)$, are parametrized as a function of $F_0$, $F_{N_{max}}$, $\lambda, \alpha$, and $\gamma$. Since the errors in $\lambda, \alpha$ and $\gamma$ have a Gaussian distribution, we can generate multiple curves (with different values of $\lambda, \alpha$ and $\gamma$ sampled from the respective Gaussians). These newly generated curves have their own intersection times, $t_{\text{int}}$. We then report the mean and $2 \sigma$ error bars for the intersection times generated this way. 

\begin{table}[!]
	\centering
	\resizebox{\columnwidth}{!}{
		\begin{tabular}{|| cccccc ||}
			\hline
			\hline
			
			DD    & $\tau/ 90 \text{ns}$ & $\lambda$       & $\alpha$    &         $\gamma$ & $t_{\mathrm{int}}$       \\ 
			\hline
			\hline
			
			Free  &                    1 & $28.9 \pm 1.2$  & $910 \pm 5$ & $0.73 \pm  0.12$ & 0               \\ 
			XY4   &                    1 & $88.4 \pm 0.3$  & $\infty$    &                0 & $108 \pm 4$ \\ 
			XY4   &                    2 & $73.5 \pm 0.7$  & $\infty$    &                0 & $95 \pm 7$  \\ 
			XY4   &                    3 & $72.3 \pm 0.5$  & $\infty$    &                0 & $99 \pm 5$  \\ 
			XY4   &                    4 & $74.8 \pm 0.4$  & $\infty$    &                0 & $100 \pm 4$ \\ 
			XY4   &                    5 & $67.0 \pm 0.6$  & $\infty$    &                0 & $91 \pm 6$  \\ 
			XY4   &                    6 & $61.6 \pm 0.7$  & $\infty$    &                0 & $85 \pm 6$  \\ 
			$(XI)^N$    &                    2 & $79.2 \pm 0.7$  & $\infty$    & $0.22 \pm  0.02$ & $89 \pm 5$  \\ 
			$(YI)^N$    &                    2 & $79.7 \pm 0.7$  & $\infty$    & $0.21 \pm  0.03$ & $89 \pm 5$  \\ 
			$(ZI)^N$    &                    2 & $63.9 \pm 1.0$  & $\infty$    & $0.29 \pm  0.03$ & $87 \pm 10$  \\ 
			GA8a  &                    1 & $78.2 \pm 0.3$  & $\infty$    &                0 & $95 \pm 3$  \\ 
			GA16a &                    1 & $95.5 \pm 0.6$ & $197.4 \pm 0.6$    &                $0.19 \pm 0.01$ & $115  \pm 3$ \\ 
			GA32a &                    1 & $88.0 \pm 0.7$  & $\infty$    &                0 & $104 \pm 8$ \\ 
			\hline
			\hline
			
		\end{tabular}
	}
	\caption{Performance summary of the different DD pulses we implemented on the IBMQX5. The XY4 sequence is the universal decoupling sequence discussed in the main text. The $X\cdots X$ and $Y\cdots Y$ sequences are specific to pure dephasing errors. The $(ZI)^N$ sequence is suppresses pure SE errors. All three of these sequences underperform the XY4 sequences, but performance is better after suppression of pure dephasing errors. The GA sequences are discussed in Appendix~\ref{app:GA}.}
	\label{tab:resultsDD}
\end{table}

\section{DD \textit{vs} free evolution correlation plots, as a function of initial state}
\label{app:initstate}

\begin{figure*}[!htb]
\raggedright
\subfigure{\hspace{-0.5cm} \includegraphics[scale=0.35]{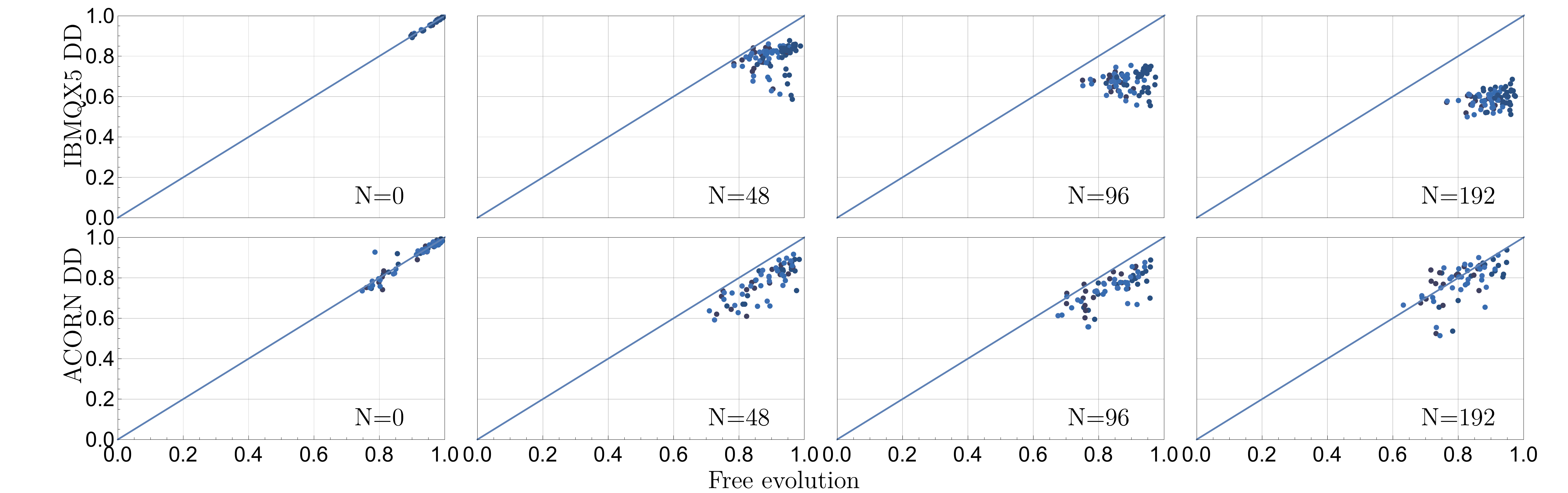}\label{fig:correlationZero0}}
\subfigure{\hspace{-0.5cm} \includegraphics[scale=0.35]{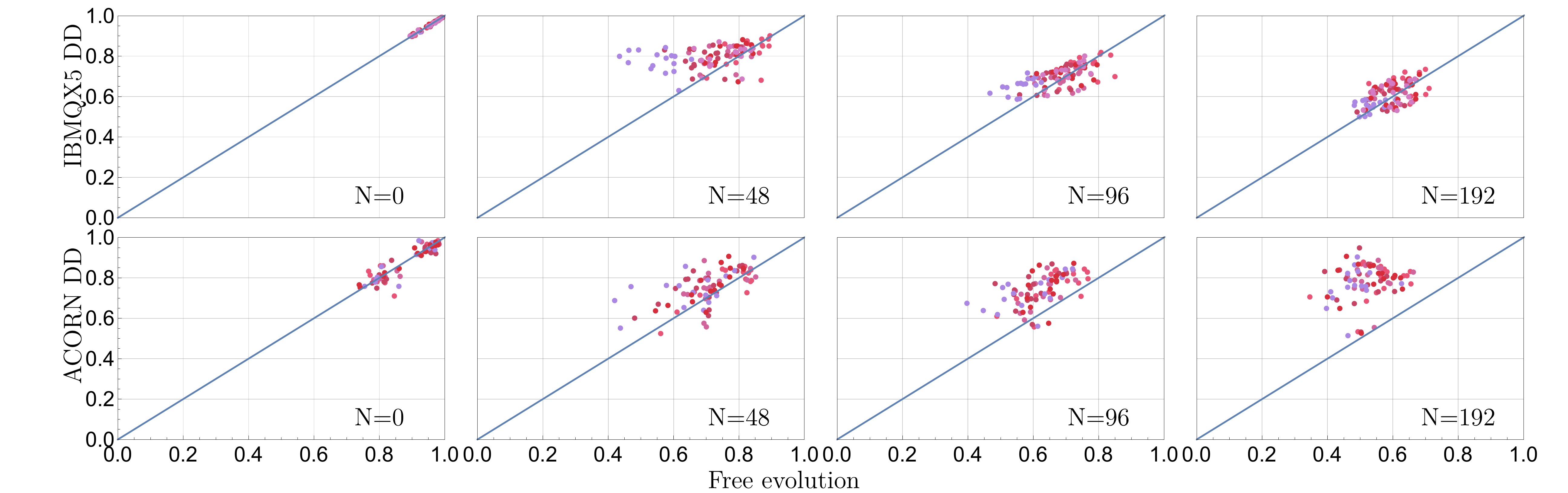}\label{fig:correlationZero1}}
\subfigure{\hspace{-0.5cm} \includegraphics[scale=0.35]{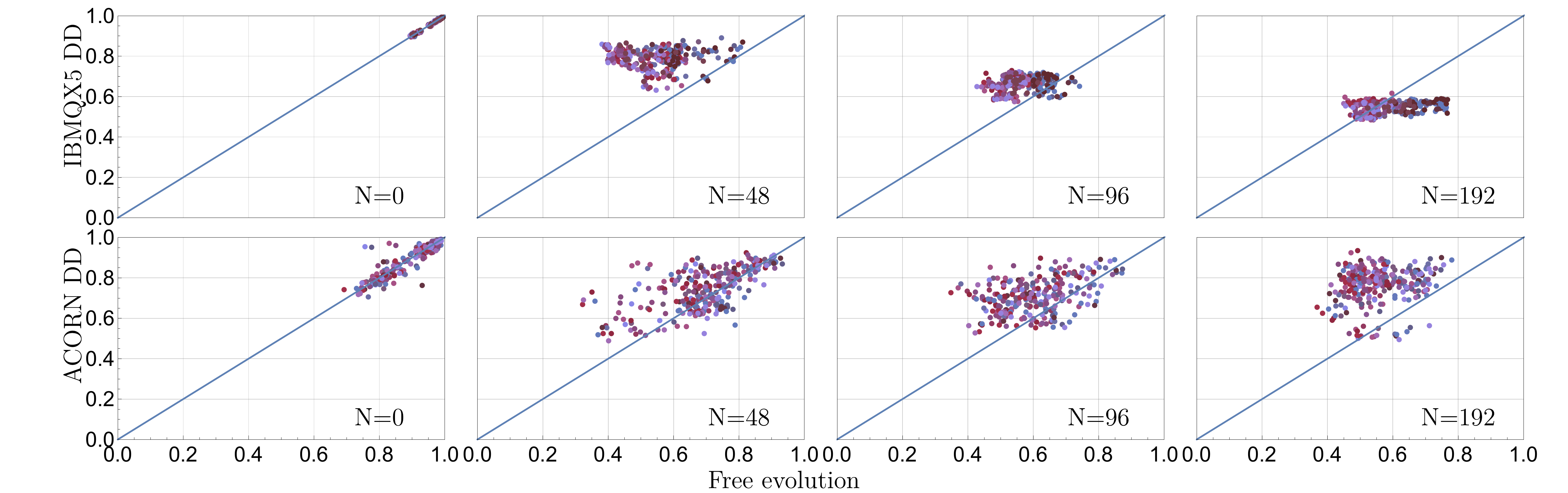}\label{fig:correlationZero2}}
\caption[ ]{(Color online) Correlation plot of the fidelities under DD compared to free evolution, for IBMQX5 and Acorn. Initial conditions are color coded as a function of $\theta$. Top panel: $\theta \in [0,\frac{\pi}{3})$; middle panel: $\theta \in [\frac{\pi}{3},\frac{2\pi}{3})$; bottom panel: $\theta \in [\frac{2\pi}{3},\pi]$. In (a) the initial state is close to the ground state $\ket{0}$ and DD is worse than free evolution. In (b) the initial state is a close to an equal superposition, thus susceptible to dephasing, and DD is overall better than free evolution, especially for Acorn. In (c) the initial state is close to the excited state $\ket{1}$ and DD is again better than free evolution at intermediate $N$ for IBMQX5, and at all $N$ for Acorn.}
\label{fig:correlationZero}
\end{figure*}

As seen in Fig.~\ref{fig:angDep}, the evolution of initial states of the form $\cos ({\theta}/{2}) \ket{0} + \sin({\theta}/{2}) \ket{1}$ varies depending on $\theta$. Overall, we find two qualitatively different behaviors depending on whether $\theta \in [0,\frac{\pi}{3})$ or $\in[\frac{\pi}{3}, \pi]$. We make this explicit by plotting the fidelity for each state under DD vs under free evolution. Each data point in Fig.~\ref{fig:correlationZero} corresponds to a single initial condition on a single qubit of the respective machines. Data points corresponding to each initial condition have been color-coded using $\theta$ with $\theta = 0$ set to blue and $\theta = \pi$ set to red. Different data points of the same color refer to fidelities for the same initial condition acquired from different qubits. Points  above the diagonal indicate an advantage for DD over free evolution.

For initial conditions that are closer to the state $\ket{0}$, we find that most of the points remain below the diagonal, indicating that performing DD in fact reduces their fidelity over time. For states that are farther from $\ket{0}$, most of the points are above the diagonal. For IBMQX5 the improvements tend to disappear as time increases, while they are retained for Acorn. Overall, free evolution preserves states close to the ground state better than DD, but both superposition states (susceptible to dephasing) and states close the excited state $\ket{1}$ (susceptible to SE) benefit from DD relative to free evolution. Overall, fidelities improve more substantially on Acorn under DD than on IBMQX5, consistent with Fig.~\ref{fig:timeEvolAvg}. 

\begin{figure}[!htb]
\raggedright
  \includegraphics[width = .48\textwidth]{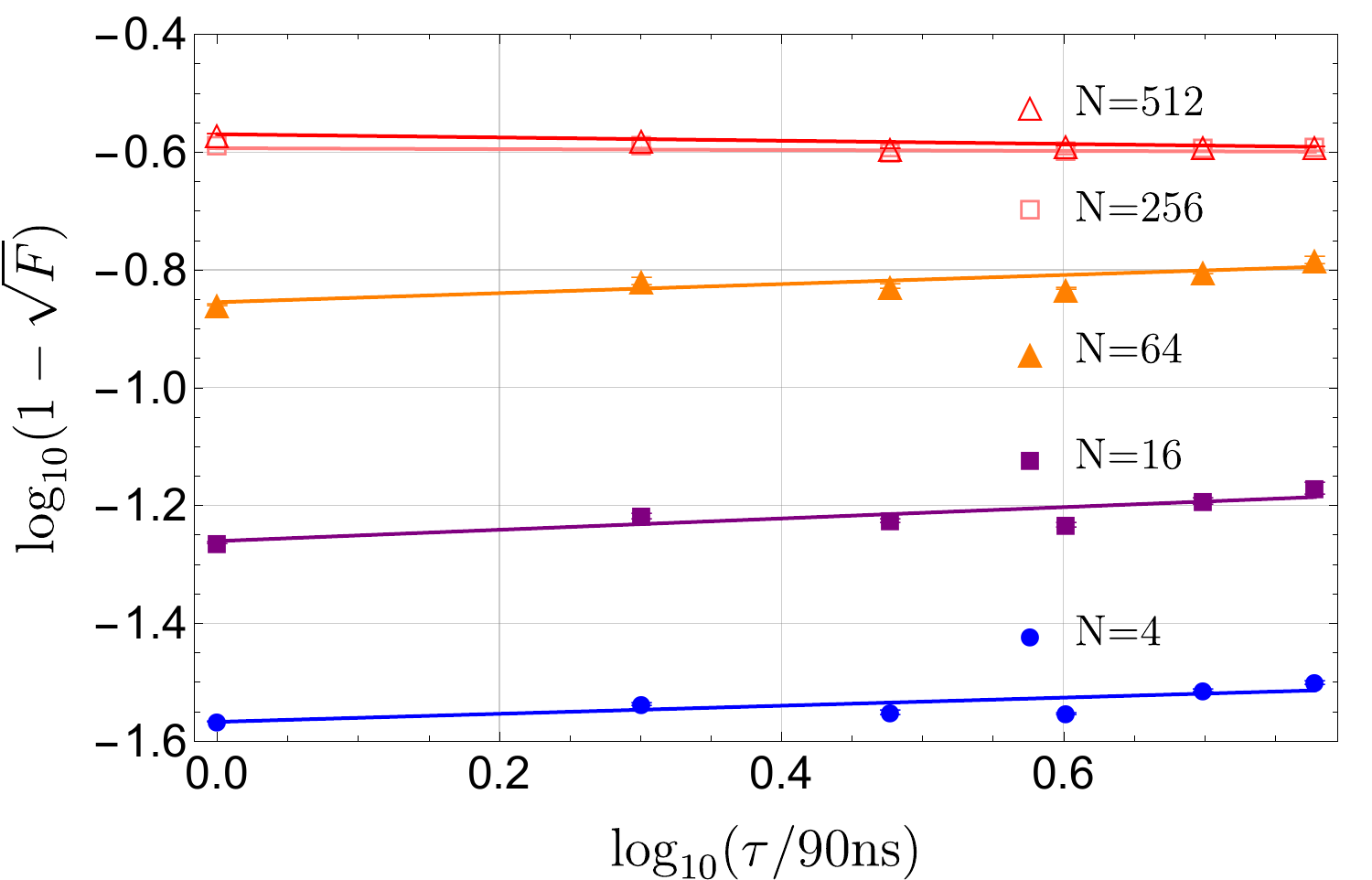}
\caption{(Color online) Infidelity scaling as a function of $\tau$ for different numbers of pulses $N$. 
}
\label{fig:infid}
\end{figure}

\begin{figure*}[!htb]
	\subfigure{\includegraphics[scale=0.38]{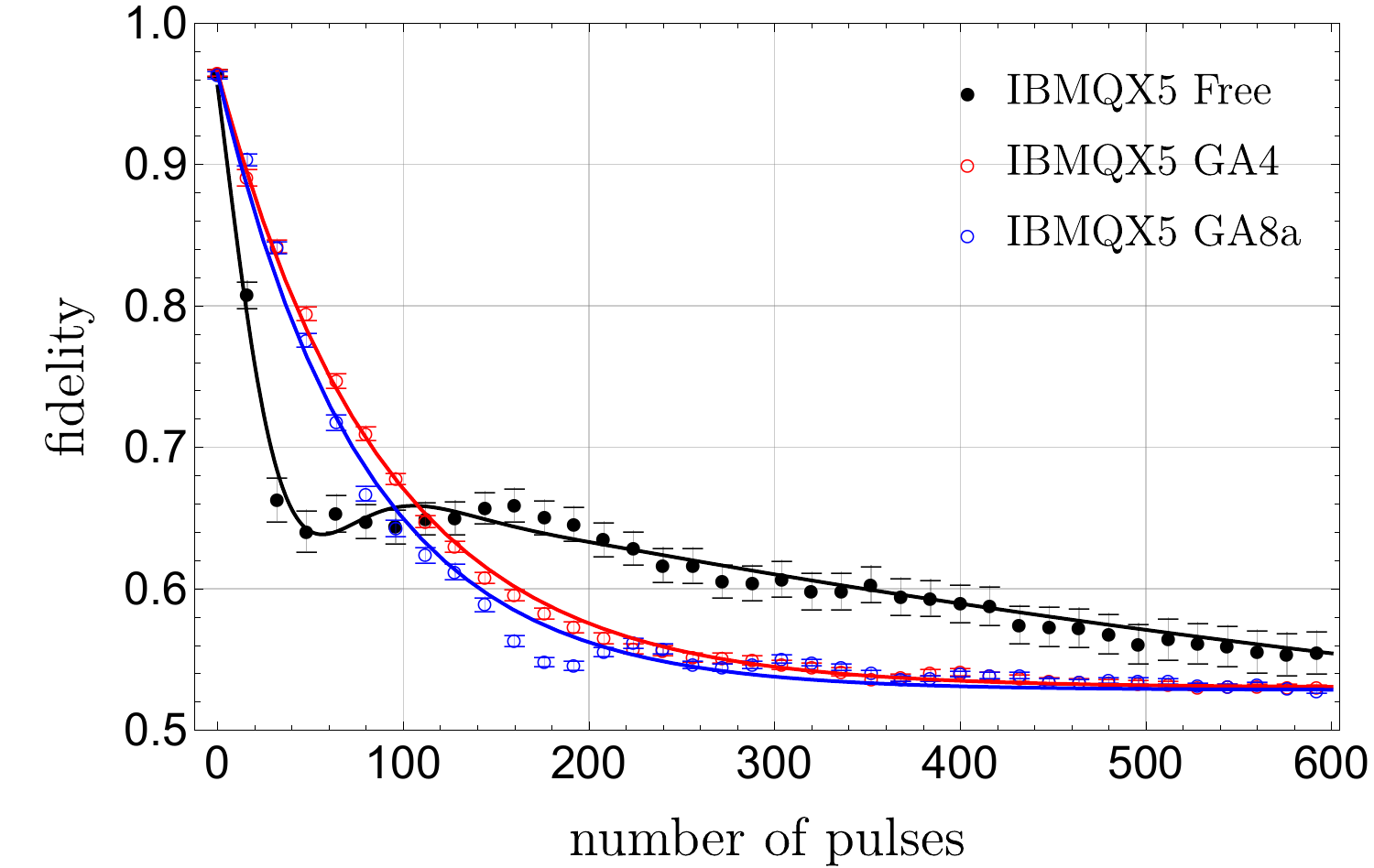}}
	\subfigure{\includegraphics[scale=0.38]{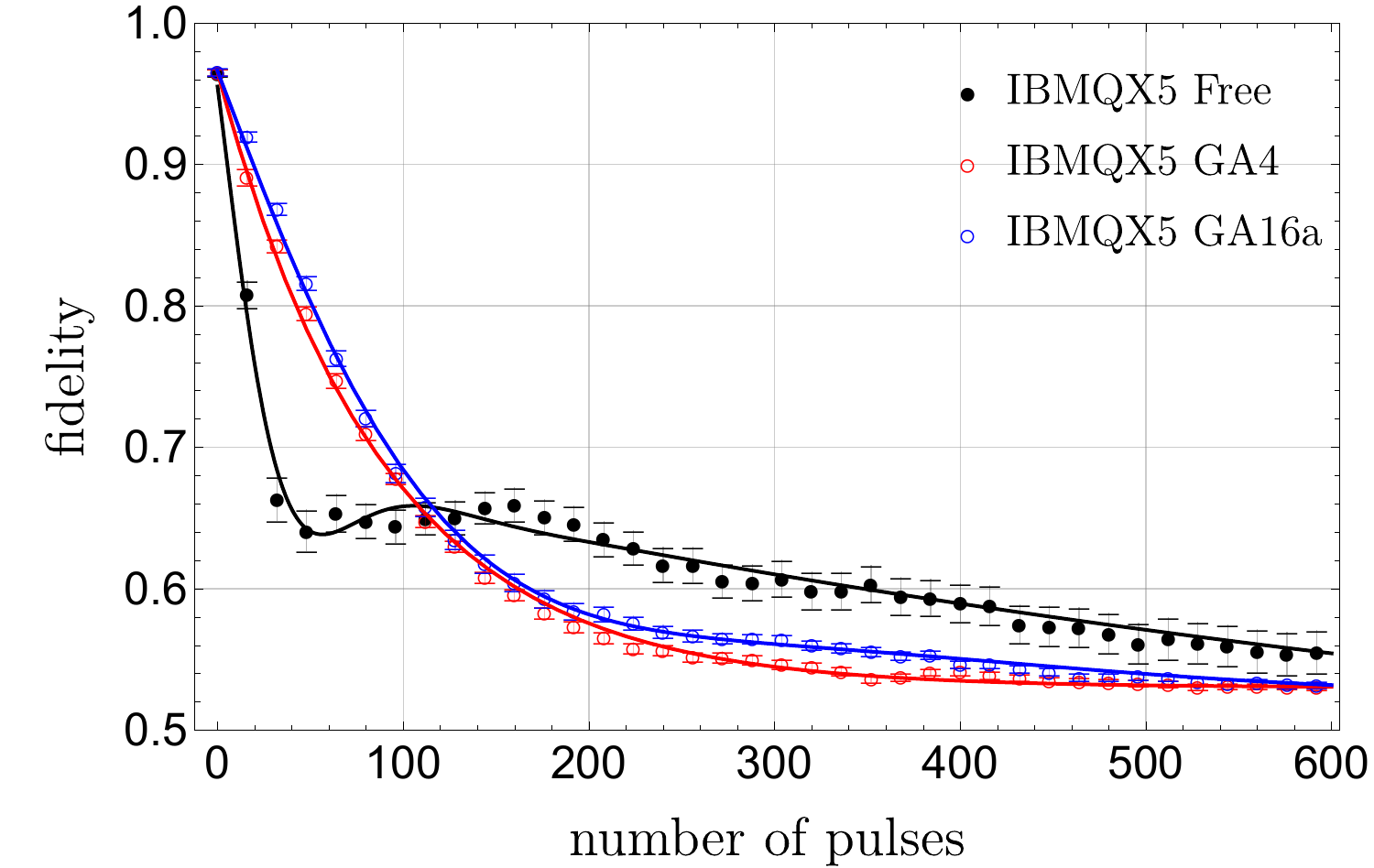}}
	\subfigure{\includegraphics[scale=0.38]{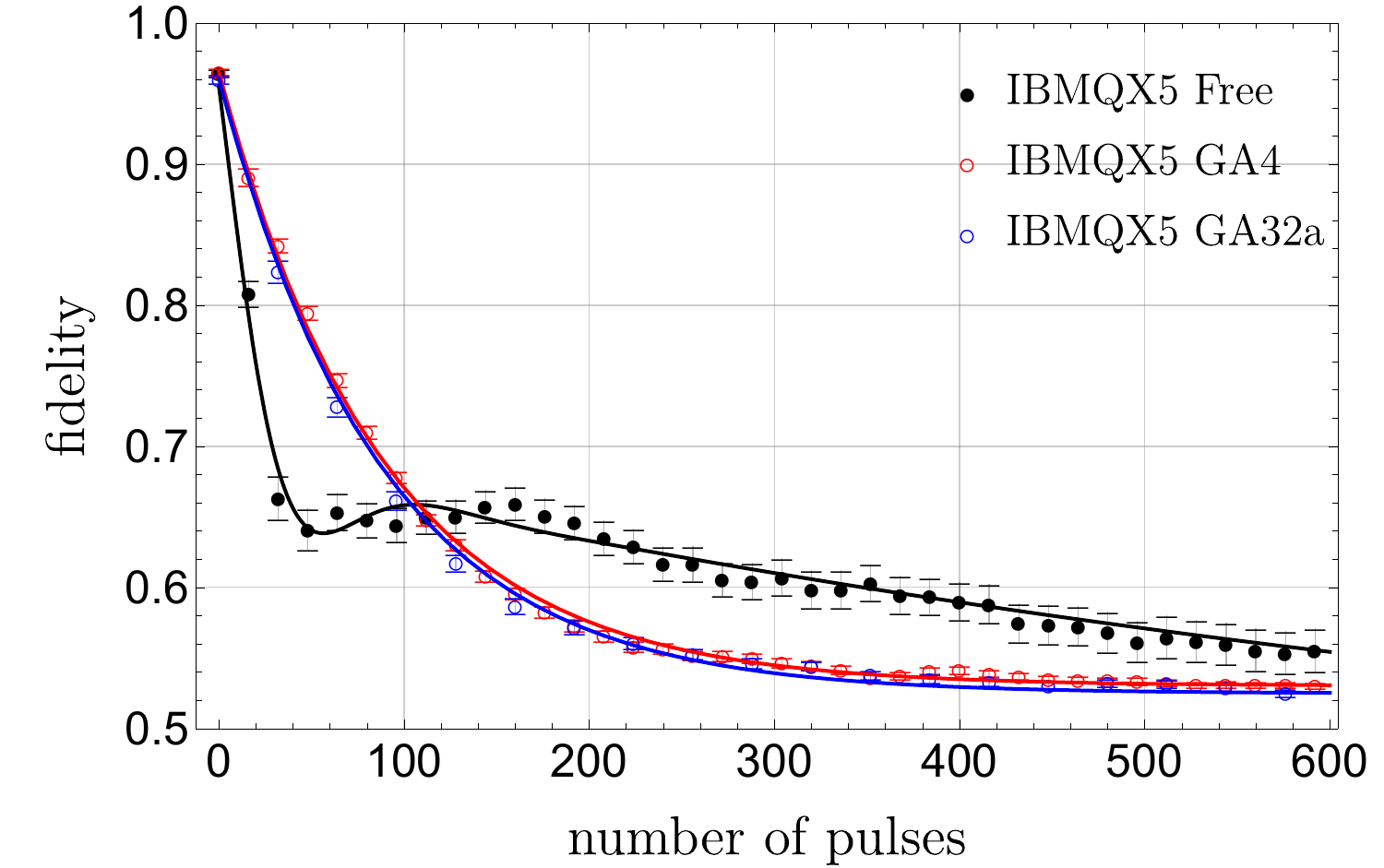}}
	\caption[ ]{(Color online) Fidelity decay results for the GA-based DD sequences compared to XY4 and free evolution, on IBMQX5. Left: GA$_{8a}$; middle: GA$_{16a}$; right: GA$_{32a}$. A small improvement over XY4 is seen for the GA$_{16a}$ sequence.}
	\label{fig:GA8-16-32}
\end{figure*}

\section{DD tailored for dephasing and spontaneous emission}
\label{app:SE}

As we have noted, both dephasing and SE play important roles in reducing the fidelity of free evolution. We thus tested sequences tailored to each of these noise sources.

To suppress pure dephasing, which results from a system-bath interaction term of the form $\sigma^z\otimes B$ (where $B$ is a bath operator), it suffices to apply the the $(XI)^N$ or $(YI)^N$ sequence, since the error $\sigma^z$ anticommutes with both $X$ and $Y$. To suppress SE, which results from a system-bath interaction term of the form $\sigma^-\otimes B$ (a consequence of the inevitable coupling of any quantum system to the vacuum electromagnetic field~\cite{Scully:book}), it suffices to apply the the $(ZI)^N$ sequence, since the error $\sigma^- = |0\rangle\langle 1| = (\sigma^x +i \sigma^y)/2$ anticommutes with $Z$. In both cases, this results in an effective evolution wherein the errors are time-reversed under the DD sequence.

The results are given in Table~\ref{tab:resultsDD}. Unsurprisingly, the performance of the specialized $(XI)^N$, $(YI)^N$, and $(ZI)^N$ sequences is worse than that of the universal XY4 sequence. However, they account for a substantial increase in the initial decay time $\lambda$, with the $\lambda$ value for the $(XI)^N$, $(YI)^N$ sequences accounting for nearly $90\%$ of the $\lambda$ value of the XY4 sequence. Table~\ref{tab:resultsDD} shows that performance is better after suppression of pure dephasing errors, which indicates that these errors dominate over SE errors.

\section{Infidelity as a function of pulse spacing}
\label{app:dist}

For ideal DD pulses the distance between the free and dynamically decoupled states is bounded as~\cite{daniel2008distance,daniel2007performance}:
\begin{equation}
     D [\rho_{S}(T), \rho_{S}^{0}(T)] \leq \frac{1}{2} \left( e^{2c T^2/N} -1 \right) \leq 2c \frac{T^2}{N} ,
\end{equation}
where $D(\rho,\sigma) = \frac{1}{2} ||\rho - \sigma ||_{1}$ is the trace-norm distance between quantum states $\rho$ and $\sigma$,
$T$ is the total evolution time, $\rho_{S}(T)$ ($\rho_{S}^{0}(T)$) is the state of the system under DD (free evolution), $N$ is the total number of (non-identity) DD pulses, and 
$c$ is a constant directly related to the operator norms of the bath and system-bath Hamiltonian. 
Since $T = \tau N$, we have for the Uhlmann fidelity:
\begin{equation}
1- \sqrt{F[\rho_{S}(T), \rho_{S}^{0}(T)]} \leq D [\rho_{S}(T), \rho_{S}^{0}(T)] \leq 2c \tau^2 N ,
\end{equation}
which leads to Eq.~\eqref{eq:infid} from the main text. 

Figure~\ref{fig:infid} shows the scaling of the infidelity. An extraction of the slopes and offsets of the straight line fits shown (with additional results for other $\tau$ and $N$ values not shown here) leads to Fig.~\ref{fig:distanceSlopeIntercept} shown in the main text.

\section{Higher order sequences based on genetic algorithms}
\label{app:GA}

The XY4 sequence we have focused on in this work can be  viewed as the first in a hierarchy of   sequences studied using genetic algorithms (GAs), shown numerically to offer increasing improvements at fixed pulse interval and pulse duration~\cite{genetic}. As far as we know, these GA-based sequences have not been tested experimentally, prior to this work. 

To describe the GA-based sequences, first let
\begin{equation}
    U(t) = P_N f_{\tau} P_{N-1} f_{\tau} ... P_2 f_{\tau} P_1 f_{\tau}, 
\end{equation}
where $P_j$ is the unitary corresponding to the $j$th pulse and $f_{\tau} = e^{-i H \tau}$ is the {free-evolution} unitary generated by the joint system-bath Hamiltonian $H$ and lasting time $\tau$.
The first few GA-based sequences are then~\cite{genetic}:
\bes
\begin{align}
\text{GA}_{4} & := P_1 f_{\tau} P_2 f_{\tau} P_1 f_{\tau} P_2 f_{\tau}  \\
\text{GA}_{8a} & := I f_{\tau} P_1 f_{\tau} P_2 f_{\tau} P_1 f_{\tau} I f_{\tau} P_1 f_{\tau} P_2 f_{\tau} P_1 f_{\tau} ,\\
\text{GA}_{16a} & := P3(\text{GA}_{8a}) P3(\text{GA}_{8a}), \\
\text{GA}_{32a} & := \text{GA}_4[\text{GA}_{8a}].
\end{align}
\ees
where $P_1, P_2$ are single-qubit Pauli operators such that $P_1 \neq P_2 \in \{X, Y, Z \}$. We set $P_1 =X, P_2 =Y$ for $\text{GA}_4$ and $P_1 =X, P_2 =Z$ for $\text{GA}_{8a}$. With this choice $\text{GA}_4 =\text{XY}_4$.

The results for all three GA-based sequences are shown in Fig.~\ref{fig:GA8-16-32}, along with XY4 and free evolution. Fit results are reported in Table~\ref{tab:resultsDD}.  We find that $\text{GA}_{32a}$ performance is similar to $\text{GA}_{4}$, while $\text{GA}_{8a}$ does slightly worse. However, $\text{GA}_{16a}$ is the best sequence we have found so far, slightly outperforming XY4.


\end{document}